\documentclass{emulateapj}

\usepackage{fancyhdr}
\usepackage{color}
\usepackage{hyperref}
\usepackage{epstopdf}

\newcommand{\myemail}{adele.plunkett@yale.edu}

\newcommand{\msun}{\mbox{M$_\odot$}}
\newcommand{\kms}{\mbox{km s$^{-1}$}}
\newcommand{\dg}{\mbox{$^\circ$}}
\newcommand{\as}{\mbox{$^{\prime\prime}$}}

\def\farcmin{\hbox{$.\!\!^{\prime}$}}

\slugcomment{Accepted for publication in ApJ}

\shorttitle{Outflows in NGC 1333}
\shortauthors{Plunkett et al. (2013)}

\begin{document}


\title{CARMA observations of protostellar outflows in NGC 1333}


\author{Adele L. Plunkett \altaffilmark{1} and H\'{e}ctor G. Arce}
\affil{Department of Astronomy, Yale University,
    P.O. Box 208101, New Haven CT 06520, USA}
%\email{\myemail}

\author{Stuartt A. Corder }
\affil{Joint ALMA Observatory, Av. Alonso de C\'{o}rdova 3107, Vitacura, Santiago, Chile} %scorder@alma.cl

\author{Diego Mardones}
\affil{Departameto de Astronom\'{i}a, Universidad de Chile, Casilla 36-D, Santiago, Chile} %diego@das.uchile.cl

\author{Anneila I. Sargent}
\affil{Astronomy Department, California Institute of Technology, 1200 East California Boulevard, Pasadena, CA 91125, USA} %afs@astro.caltech.edu

\and

\author{Scott L. Schnee}
\affil{National Radio Astronomy Observatory, 520 Edgemont Road, Charlottesville, VA 22903, USA} %sschnee@nrao.edu


\altaffiltext{1}{\myemail, NSF Graduate Research Fellow}


\begin{abstract}
We present observations of outflows in the star-forming region NGC 1333 using the Combined Array for Research in Millimeter-Wave Astronomy (CARMA).  We combined the $^{12}$CO and  $^{13}$CO (1-0) CARMA mosaics with data from the 14-m Five College Radio Astronomy Observatory (FCRAO) to probe the central, most dense and active region of this protostellar cluster at scales from 5$^{\prime\prime}$ to $7^\prime$ (or 1000 AU to 0.5 pc at a distance of 235 pc).  We map and identify $^{12}$CO outflows, and along with $^{13}$CO data we estimate their mass, momentum and energy.  Within the $7^\prime\times7^\prime$ map, the 5$^{\prime\prime}$ resolution allows for a detailed study of morphology and kinematics of outflows and outflow candidates, some of which were previously confused with other outflow emission in the region.  In total, we identify 22 outflow lobes, as well as 9 dense circumstellar envelopes marked by continuum emission, of which 6 drive outflows.  We calculate a total outflow mass, momentum and energy within the mapped region of  6 \msun, 19 \msun km s$^{-1}$,  and 7 $\times10^{44}$ erg, respectively.   Within this same region, we compare outflow kinematics with turbulence and gravitational energy, and we suggest that outflows are likely important agents for the maintenance of turbulence in this region.  In the earliest stages of star formation, outflows do not yet contribute enough energy to totally disrupt the clustered region where most star formation is happening, but have the potential to do so as the protostellar sources evolve. Our results can be used to constrain outflow properties, such as outflow strength, in numerical simulations of outflow-driven turbulence in clusters.
\end{abstract}





%%%%%%%%%%%%%%%%%%%%%%%%%%%%%%%%%%%%%%%%%%%
\section{Introduction}

Outflows are generally understood as a necessary component of the star formation process \citep[e.g.][]{Shu87}.  They expel mass, remove the excess angular momentum accumulated during gravitational infall and allow accretion onto a forming protostar.  Not only does an outflow play an important role in the formation of the protostar which drives it, an outflow also likely impacts its surrounding environment, and in dense cluster environments where many protostars are driving outflows, the fate of the cluster may depend on the level of outflow activity.  For example, energetic outflows inject momentum and energy into the cloud, and in the process they may disperse the surrounding gas and feed turbulent motions \citep{Arc07}. Given that the majority of stars form within embedded clusters with many members \citep{Lad03}, understanding these environments is vital to understanding the formation of most stars. 

Particularly in low-mass star-forming regions, outflows may disrupt the surrounding environment enough to limit the lifetime of their parent molecular cloud \citep{Har01}.  Recent analytical and numerical studies have shown that outflows can interact with the cloud and drive turbulent motions very efficiently \citep{Mat07,Nak07,Cun09,Car09}.  Other numerical studies suggest that outflows disrupt dense clumps and affect cloud structure, but are inefficient at driving turbulence within a cloud \citep{Ban07}.  The question remains as to whether outflows have enough energy and momentum to alter the surrounding molecular gas and trigger the formation of stars (as indicated by the models of \citealt{Fos96}), or possibly disperse the cluster gas \citep[e.g.][]{Ben04,Arc10}.   While the underlying physics of outflow-cloud interactions can be simulated, observations sensitive to a range of spatial scales to probe distinct outflows and the surrounding cloud are critical for constraining characteristics of protostellar clusters, and specifically numerical simulations that model outflow-induced turbulence.

Here we present observations of CO in the low-mass star-forming region NGC 1333 with the primary goal to identify and characterize outflow activity.  We also present continuum observations made simultaneously with the CO observations, and although the continuum data were not the priority of our observing plan, they allow a qualitative and quantitative study of several outflow-driving sources in the region.  In total, we identify twenty-two outflow lobes, as well as nine dense circumstellar envelopes marked by continuum emission within an $\sim0.23$ parsec$^2$ area.  Our observations probe physical scales over two orders of magnitude and outflow velocities up to $\pm10$ \kms \ from the cloud velocity, which we find necessary to study the distinct morphologies and kinematics of previously known outflows as well as new outflow candidates that until now were confused with other outflow emission in the region.  We measure mass, momentum and energy of individual outflows, as well as mean characteristics of outflows in our map, providing important parameters for numerical simulations of clustered star formation.  We also compare outflow energetics with turbulence and gravity in the region.  We suggest that outflows have the potential to maintain turbulence in this region, and eventually may drive enough energy to contribute towards disruption of the central, clustered region where most star formation is happening.  NGC 1333 is the first in a survey of star-forming regions that we plan to study ranging in mass and evolutionary stage, in order to investigate the impact of outflows on their surrounding environments, given environments with a range of characteristics.

\subsection{Description of the Region: NGC 1333}

\object{NGC 1333}, a reflection nebula on the near surface of the L1450 dark cloud \citep{Lyn62}, is considered by many to be the prototypical, nearby cluster-forming region \citep[e.g.][and references therein]{Wal08,Pad09}.  It is the most active region of low-mass star formation in the Perseus molecular cloud complex and the nearest large-membership ($>100$) young cluster.  A variety of works have contributed to identifying and characterizing the young stellar object (YSO) members of this cluster. \citet{Str76}\footnote{Throughout we give specific source labels from Strom, Vrba \& Strom (1976) \nocite{Str76} as SVS, rather than SSV as was the convention followed by some other works following \citet{Her83}.} presented NGC 1333 as part of a survey mapping infrared dark clouds in IR at 2.2 $\mu$m ($K$ band), in addition to photometry at $J$, $H$ and $L$ bands for identified sources within the cloud.  Later \citet{Jen87} made IRAS observations at 50 and 100 $\mu$m and revealed nine distinct sources in NGC 1333, five of which are within our mapped region.  Based on near-IR photometry, \citet{Asp94} suggested that the PMS sources in NGC 1333 are clustered around SVS 13, where many studies of clustered star formation have since targeted their efforts.  Infrared excess is found in $61\%$ of PMS sources, and \citet{Lad96} conclude that the cluster has an age of $1-2\times10^6$ years, or less.  Further, given the current stellar mass ($45$ \msun) in NGC 1333, they suggest that if star formation continues at the current rate ($4.5\times10^{-5}$\ $\msun \textrm{ yr}^{-1}$) for the next five million years, NGC 1333 could develop into a very rich cluster, reminiscent of IC 348.  

Counterparts of the IR and near-IR sources throughout NGC 1333 were detected in radio with the VLA \citep{Sne86,Rod97,Rod99}; in sub-mm with SCUBA/JCMT \citep{San01}; and in sub-arcsec mm observations with BIMA \citep{Loo00}.  For a more thorough review of the region, we refer the reader to \citet{Wal08}.   

Within NGC 1333 are a large number of molecular outflows, Herbig-Haro (HH) objects, and H$_2$ knots \citep[e.g.][]{Bal96,San01,Dav08}.  Based on these indicators, we consider this cluster ideal for studying outflow-cloud interactions.  To date, molecular outflows in NGC 1333 have either been (a) observed individually with high-resolution, or (b) mapped across the extended region with lower-resolution.  Here we present for the first time high-resolution observations that also capture the complex web of outflow activity across the central $\sim7^\prime\times7^\prime$ region of NGC 1333.

\begin{figure*}[!ht]
\includegraphics[width=\linewidth,angle=0]{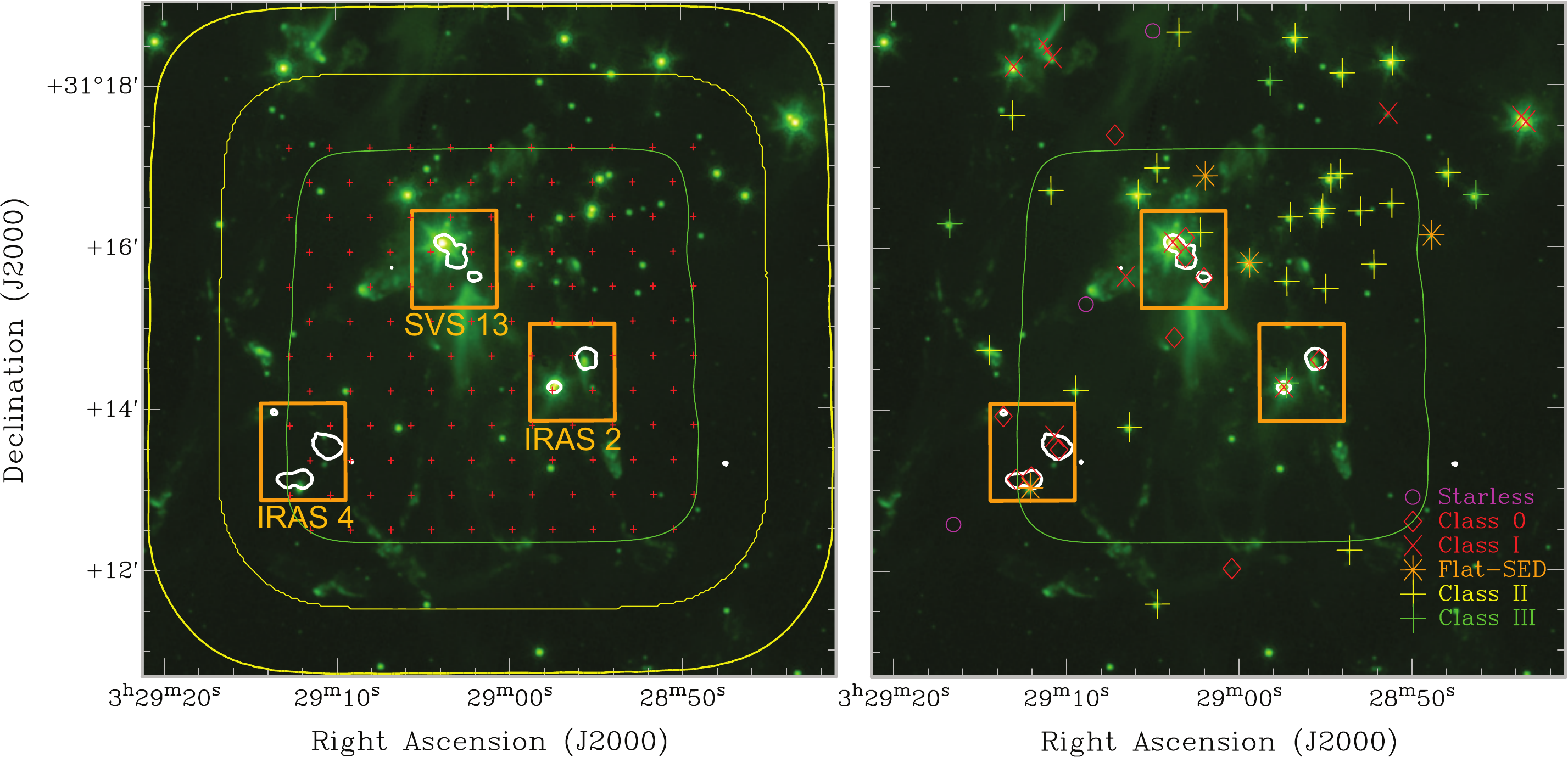}
\caption{\textit{Spitzer} IRAC 4.5 $\mu$m image of NGC 1333 \citep{Gut08}, showing the region we mapped.  More details about our observations are given in Table \ref{tab:carmamaps}. \textit{(left)} Red crosses show the 126 mosaic pointings in a hexagonal-packed pattern that comprise our CARMA maps (see \S \ref{sec:carmaobs}). Green contours indicate the region with constant sensitivity in our CARMA mosaic (see Table \ref{tab:carmamaps} for rms values), while sensitivity is degraded at the edges beyond the mosaic pattern.  The yellow (outermost) contours mark the outer edge of the mapped region, where sensitivity (between the yellow contours) is 12 mJy beam$^{-1}$, 1.1 Jy beam$^{-1}$ and 0.7 Jy beam$^{-1}$ for continuum, $^{12}$CO and $^{13}$CO, respectively.  Orange boxes indicate the regions which are shown in 2.7 mm continuum in Figure \ref{fig:contmap2}, with white contours marking 4 $\sigma$ continuum emission (i.e. the lowest-level contour in Figure \ref{fig:contmap2}).  \textit{(right)} YSOs within the region we mapped, identified in Table \ref{tab:ysos}.  White contours and orange boxes are the same as in the left panel.  The green contour is the same as the inner green contour in the left panel, indicating the region we mapped with greatest sensitivity.   }  
\label{fig:gut08}
\end{figure*}

Figure \ref{fig:gut08} shows the \textit{Spitzer} IRAC $4.5\mu$m image of NGC 1333 \citep{Gut08}, making evident the intricate outflow activity particularly in the central region.  We focus on the central $\sim7^\prime\times7^\prime$ ($\sim0.23$ pc$^2$ at 235 pc) region of NGC 1333 where 55 YSOs and 3 starless cores are found (see Table \ref{tab:ysos}), at a density of about 250 YSOs pc$^{-2}$.  This region, approximately centered on SVS 13, has one of the highest protostellar densities in the entire NGC 1333 region and overall Perseus cloud, which have average densities of about 34 YSO pc$^{-2}$ and 6 YSO pc$^{-2}$, respectively \citep{Jor06}.   

In Figure \ref{fig:gut08} and Table \ref{tab:ysos} we show and list the previously identified YSOs within the region we mapped, in an attempt to synthesize the vast literature related to this region and their subsequent naming conventions. In the table we include references which present survey-like observations of multiple sources within the region that we observed, as well as relevant naming conventions which aid the discussion of this paper and cross-references with the previous literature.

Several distances have been reported for this region, ranging from 220 pc \citep{Cer90} to 350 pc \citep{Her83}.  Here we adopt a distance of $235\pm18$ pc, which was determined by \citet{Hir08} based on parallax measurements of the H$_2$O maser in NGC 1333 SVS 13. This distance is also consistent with the distance of 250 pc assumed in other studies of the region \citep[e.g.][]{Eno06,Eva09,Cur10a,Arc11}. In a Perseus-wide study of sub-mm cores, \citet{Hat05} assumed a distance of 320 pc, based on the \textit{Hipparcos} distance to clusters within the Perseus OB2 association \citep{deZ99}. However, \citet{Eno06} cite evidence that NGC 1333 may lie in the foreground of the more distant Perseus OB2 association, or that the Perseus cloud may span a range of distances.  

%%%%%%%%%%%%%%%%%%
%fig1
%%%%%%%%%%%%%%%%%%

%%%%%%%%%%%%%%%%%%
%tab2
%table of ysos
\begin{deluxetable*}{lcccl}
%\rotate
\tabletypesize{\tiny}
%\tablewidth{9in}
\tablecaption{YSOs and starless cores within the mapped region\label{tab:ysos}}
\tablehead{\colhead{Source}&\colhead{$\alpha$(J2000)}&\colhead{$\delta$(J2000)}&\colhead{Class\tablenotemark{a}}&\colhead{Other names\tablenotemark{b}}}
\startdata
Classes S, 0, I	&		&		&	&		\\
\hline
IRAS 5	&	03:28:43.27	&	31:17:33.10	&	I	&	SVS	9, ASR 126, [GMM2008] 18, J032843.28+311732.9	\\
ASR 127	&	03:28:43.56	&	31:17:36.50	&	I	& 	[GMM2008] 45, Bolo 31(I)\tablenotemark{c} \\
ASR 41	&	03:28:51.27	&	31:17:39.50	&	I	& 	[GMM2008] 20, J032851.26+311739.3\tablenotemark{e}	\\
IRAS 2A	&	03:28:55.30	&	31:14:36.40	&	0\tablenotemark{d}	& 	VLA 7, SK 8, HRF 44(0), [GMM2008] 21(I), Bolo 38(0)\tablenotemark{c},\\
&&&& J032855.55+311436.7(0)	\\
IRAS 2B	&	03:28:57.37	&	31:14:16.20	&	I\tablenotemark{d}	&	SVS	19\tablenotemark{f}, VLA 10, SK 7, [GMM2008] 3(0), Bolo 38(I)\tablenotemark{c},	\\
&&&& J032857.36+311415.9(I)	\\
SK 1	&	03:29:00.40	&	31:12:01.50	&	0	&   	HRF 65, Bolo 41, [GMM2008] 4, J032900.55+311200.8	\\
SVS 13C	&	03:29:01.97	&	31:15:37.40	&	0	&  	MMS3, VLA 2, SK 11, H$_2$O(B)	 \\
VLA 3	&	03:29:03.00	&	31:16:02.00	&	0	&  	\\
SVS 13B	&	03:29:03.05	&	31:15:52.50	&	0	&  	MMS2, VLA 17, SK 12	 \\
SVS 13A	&	03:29:03.20	&	31:15:59.00	&	I	&	SVS	13, IRAS 3, ASR 1, MMS1, VLA 4, SK 13, HRF 43, [GMM2008] 29, Bolo 43,	\\
&&&&  J032903.78+311603.8	\\
SK 14\tablenotemark{g}	&	03:29:03.70	&	31:14:53.10	&	0	&  	VLA 19, HRF 52, [GMM2008] 5, Bolo 46, J032904.06+311446.5	\\
Bolo 44	&	03:29:04.90	&	31:18:41.20	&	S	&    		 \\
SK 15	&	03:29:06.50	&	31:15:38.60	&	I	&   	HRF 50	 \\
SK 18	&	03:29:07.10	&	31:17:23.70	&	0	&   	HRF 62	 \\
SK 16	&	03:29:08.80	&	31:15:18.10	&	S	&   	HRF 51	 \\
IRAS 4A	&	03:29:10.40	&	31:13:30.00	&	0	& 	VLA 25, SK 4, HRF 41, [GMM2008] 6, Bolo 48\tablenotemark{c}, J032910.49+311331.0	\\
J032910.65+311340.0	&	03:29:10.65	&	31:13:40.00	&	I	&     	\\
SK 20	&	03:29:10.71	&	31:18:21.20	&	I\tablenotemark{d}	&   	Bolo 49(0)\tablenotemark{c}, J032910.68+311820.6(I)	\\
IRAS 7	&	03:29:11.00	&	31:18:27.40	&	I\tablenotemark{d}	& 	HRF 46(0), [GMM2008] 35(I), J032910.99+311826.0(I)	\\
SK 21	&	03:29:11.25	&	31:18:31.70	&	I\tablenotemark{d}	& 	ASR 33, VLA 27, [GMM2008] 7(0), Bolo 49(0)\tablenotemark{c}, J032911.26+311831.4(I)	\\
IRAS 4B	&	03:29:12.00	&	31:13:10.00	&	0	&  	VLA 28, SK 3, HRF 42, [GMM2008] 8, Bolo 48\tablenotemark{c}, J032912.06+311305.4	\\
IRAS 4C	&	03:29:12.88	&	31:13:08.20	&	0	&   	SK 2	 \\
ASR 30	&	03:29:12.95	&	31:18:14.60	&	I	& 	[GMM2008] 36, Bolo 49(I)\tablenotemark{c}, J032912.97+311814.3	\\
IRAS 4D	&	03:29:13.60	&	31:13:55.00	&	0	&  	VLA 29, SK 5, HRF 48, [GMM2008] 9, Bolo 48\tablenotemark{c}, J032913.54+311358.2	\\
HRF 59	&	03:29:16.50	&	31:12:34.60	&	S	&    	 \\
\hline
Classes F, II, III	&		&		&	&		\\
\hline
J032846.21+311638.4 &	03:28:46.19	&	31:16:38.70	&	III	&	SVS	21, ASR 128, [GMM2008] 47\\
J032847.84+311655.1 &	03:28:47.82	&	31:16:55.30	&	II 	&	SVS	17, ASR 111, [GMM2008] 49\\
ASR 67	&	03:28:48.76	&	31:16:08.90	&	F\tablenotemark{d}	& 	[GMM2008] 19(I), J032847.84+311655.1(F)	\\
J032851.03+311818.5 &	03:28:51.02	&	31:18:18.50	&	II 	&	SVS	10, ASR 122, [GMM2008] 50\\
J032851.08+311632.4 &	03:28:51.07	&	31:16:32.60	&	II 	& 	ASR 44, [GMM2008] 51	\\
J032852.15+311547.1 &	03:28:52.13	&	31:15:47.20	&	II 	& 	ASR 45, [GMM2008] 53\\
J032852.92+311626.4 &	03:28:52.90	&	31:16:26.60	&	II 	& 	ASR 46, [GMM2008] 55\\
ASR 97	&	03:28:53.58	&	31:12:14.70	&	II 	& [GMM2008] 56 \\
%SK 9	&	03:28:53.90	&	31:14:53.50	&	?	&   	 \\
J032853.96+311809.3 &	03:28:53.93	&	31:18:09.30	&	II	& 	ASR 40, [GMM2008] 57 	\\
J032854.09+311654.2 &	03:28:54.07	&	31:16:54.50	&	II 	&	SVS	18\tablenotemark{h}, ASR 42, [GMM2008] 58 	\\
J032854.63+311651.1 &	03:28:54.61	&	31:16:51.30	&	II 	&	SVS	18\tablenotemark{h}, ASR 43, [GMM2008] 59 	\\
ASR 109	&	03:28:54.92	&	31:15:29.20	&	II	& [GMM2008] 60\\
J032855.08+311628.7 &	03:28:55.07	&	31:16:28.80	&	II 	& 	ASR 107, [GMM2008] 61 \\
ASR 108	&	03:28:55.15	&	31:16:24.80	&	II	& [GMM2008] 62 \\
J032856.65+311835.5 &	03:28:56.64	&	31:18:35.70	&	II 	&	SVS	11, ASR 120, Bolo 39, [GMM2008] 65 	\\
J032856.97+311622.3 &	03:28:56.95	&	31:16:22.30	&	II 	&	SVS	15, ASR 118, [GMM2008] 67 	\\
J032857.18+311534.6 &	03:28:57.17	&	31:15:34.60	&	II 	& 	ASR 17, [GMM2008] 68 	\\
J032857.21+311419.1\tablenotemark{i}	&	03:28:57.21	&	31:14:19.10	&	III	&     		\\
J032858.11+311803.7 &	03:28:58.11	&	31:18:03.70	&	III	& 	ASR 36	\\
J032859.32+311548.7 &	03:28:59.32	&	31:15:48.50	&	F	&	SVS 16, ASR 106, [GMM2008] 73	\\
%SK 10	&	03:29:00.80	&	31:14:24.40	&	?	&   	 \\
J032901.88+311653.2 &	03:29:01.88	&	31:16:53.20	&	F	& 	ASR 11 	\\
ASR 3	&	03:29:02.16	&	31:16:11.40	&	II	& [GMM2008] 76\\
ASR 63	&	03:29:03.39	&	31:18:40.10	&	II	&  [GMM2008] 80\\
J032904.68+311659.0 &	03:29:04.67	&	31:16:59.20	&	II 	& 	ASR 105, [GMM2008] 84 	\\
J032904.73+311134.9 &	03:29:04.73	&	31:11:35.00	&	II 	& 	ASR 99, [GMM2008] 85 	\\
J032905.78+311639.6 &	03:29:05.76	&	31:16:39.70	&	II 	&	SVS	14, ASR 7, VLA 22, [GMM2008] 88	\\
J032906.33+311346.4 &	03:29:06.32	&	31:13:46.50	&	II 	& 	ASR 53, [GMM2008] 89 \\
%SK 17	&	03:29:08.78	&	31:17:59.50	&	?	&   	 \\
J032909.40+311413.8 &	03:29:09.41	&	31:14:14.10	&	II 	& 	ASR 54, [GMM2008]	 135\\
J032910.84+311642.6 &	03:29:10.82	&	31:16:42.70	&	II 	& 	ASR 23, [GMM2008] 99 \\
J032912.06+311301.7\tablenotemark{j}	&	03:29:12.06	&	31:13:01.70	&	F	&     		\\
ASR 28	&	03:29:13.04	&	31:17:38.40	&	II	&  [GMM2008] 105 \\
J032914.40+311444.1 &	03:29:14.40	&	31:14:44.10	&	II 	&     		\\
J032916.69+311618.2 &	03:29:16.69	&	31:16:18.20	&	III	&     		\\
\enddata
\tablenotetext{a}{Classifications as starless (S), Class 0, Class I, flat-SED (F), Class II or Class III, where information is available in the literature.   }
\tablenotetext{b}{Prefixes and associated references: ``SVS'' \citet{Str76}; ``IRAS'' \citet{Jen87}; ``ASR'' \citet{Asp94}; ``MMS'' \citet{Chi97}; ``VLA'' \citet{Rod97,Rod99}; ``SK'' \citet{San01}; ``HRF'' \citet{Hat07a}; [GMM2008] \citet{Gut08}; ``Bolo'' \citet{Eno09}; ``J'' \citet{Eva09} (preceded by ``SSTc2d'')}
\tablenotetext{c}{More than one protostar associated with a Bolocam source, or source is in a crowded region.}
\tablenotetext{d}{Source classification differs among references.  In these cases, we assume the classification from \citet{Eva09}, but we give the classifications for each reference, where appropriate.}
\tablenotetext{e}{Classified as Class I based on $\alpha$(=0.62), but with high $T_{Bol}$(=1000 K). }
\tablenotetext{f}{Source position is coincident, but \citet{Rod99} suggest this is the G2 IV star BD +30\dg547 (=ASR 130)}
\tablenotetext{g}{Sources from different references within about 8\arcsec.}
\tablenotetext{h}{Source from \citet{Str76} is very near to two identified sources such that we cannot distinguish to which it corresponds.}
\tablenotetext{i}{Near IRAS 2B}
\tablenotetext{j}{Near IRAS 4C}

\end{deluxetable*}
\clearpage
%%%%%%%%%%%%%%%%%%

%%%%%%%%%%%%%%%%%%%%%%%%%%%%%%%%%%%%%%%%%%%%%%
\section{Observations and Data} \label{sec:obs}

\subsection{CARMA Observations} \label{sec:carmaobs}

We observed with CARMA in the D- and E-array configurations, mapping a $8.4\arcmin\times8.2\arcmin$ region of NGC 1333, centered at RA=03$^{\textrm{h}}29^{\textrm{m}}01^{\textrm{s}}$, Dec=+31\dg15$\arcmin$00\as. In total, we observed 41.6 hours with the D-array configuration and 15.7 hours with the E-array configuration. Combining the D- and E-array configuration observations produced a dataset with baselines ranging from 8.5 to 148 m.  CARMA is a heterogeneous array which, at the time of observations in 2008 October and 2009 March, consisted of nine antennas with diameters of 6.1 m and six antenna with diameters of 10.2 m (with half-power beam widths of 100\as\ and 60\as\ at 115 GHz, respectively).  Our mosaic consisted of 126 pointings in a hexagonal-packed pattern, with horizontal spacing of 30\arcsec\ and vertical spacing of 25.8\arcsec.  This provides better-than Nyquist sampling considering the 60\as\ HPBW of the 10.2 m antennas.  The resulting synthesized beams and the rms of the maps are given in Table \ref{tab:carmamaps}.  

%%%%%%%%%%%%%%%%%%
%tab1
\begin{deluxetable*}{lccccccccc}
\tabletypesize{\tiny}
\setlength{\tabcolsep}{0.04in} 
\tablecaption{CARMA Maps Summary \label{tab:carmamaps}}

\tablehead{\colhead{Line}  & \colhead{Rest Frequency}  & \multicolumn{2}{c}{HPBW} & \colhead{Beam PA} & \colhead{$v_{min}$} & \colhead{$v_{max}$} & \colhead{Channel width} &\colhead{Bandwidth}& \colhead{RMS} \\ 
\cline{3-4} 
\colhead{}&\colhead{(GHz)}  &  \colhead{maj($^{\prime\prime}$)} & \colhead{min ($^{\prime\prime}$)} & \colhead{($^\circ$)} & \colhead{(km s$^{-1}$)}& \colhead{(km s$^{-1}$)} & \colhead{(km s$^{-1}$)} & \colhead{(MHz)}& \colhead{(Jy beam$^{-1}$)} } 
\startdata
$^{12}$CO   &  115.27       & 5.7 &4.7  & 82.7 & -2.3 & 17.6 & 0.32 & 7.69 & 0.2\\
$^{13}$CO   &  110.20       & 5.9 &5.0  & 72.9 & -0.9 & 17.6 & 0.33 & 7.69 & 0.1\\
Continuum & 112.7 &  6.5&  5.3 &  84.0& \nodata & \nodata  &  \nodata & 1531  & 0.002\\
\enddata
\end{deluxetable*}
%%%%%%%%%%%%%%%%%%

We simultaneously observed the $J=1-0$ transitions of $^{12}$CO (115.27 GHz) in the upper side band (USB), and $^{13}$CO (110.20 GHz) and C$^{18}$O (109.78 GHz) in the lower side band (LSB). The spectral windows of $^{12}$CO and $^{13}$CO had widths of 8 MHz, and the spectral window of C$^{18}$O had width of 2 MHz.  With 63 channels in each window, the velocity resolutions were 0.317 \kms , 0.332 \kms, and 0.083 \kms\ for $^{12}$CO, $^{13}$CO, and C$^{18}$O, respectively.  In Table \ref{tab:carmamaps} we summarize the $^{12}$CO, $^{13}$CO and continuum maps, including the spectral set-up and the corresponding minimum and maximum velocity of the lines we present here.  We note that C$^{18}$O data are not included in the current paper.  Our assumed cloud velocity of $v_{cloud}=8$ \kms\ (see \S \ref{sec:molecules}) corresponds to channels 33 and 28 (of 63 channels total) in the $^{12}$CO and $^{13}$CO data cubes, respectively.  Hence, we detect linewings up to about $\pm 10$ \kms\ from the cloud velocity, and although the limited velocity range will result in missing the very high-velocity emission of a few outflows, we chose the spectral set-up that would give us a sufficient combination of both spectral resolution and coverage.

Our correlator setup included one 500 MHz window, with data taken in both the USB and LSB for a total of 1 GHz.  The central frequencies for the wide-band windows in the LSB and USB observed with the D-array configuration were 109.9 GHz and 115.5 GHz, respectively, and the central frequencies for the windows observed with the E-array configuration were 110.6 GHz and 114.9 GHz, respectively.  The wide-band windows observed with the D- and E-array configurations did not overlap in frequency coverage, resulting in a total wide-band coverage of 2 GHz, including LSB and USB windows from the D- and E-array configuration observations.  We omitted eleven D-array configuration channels which showed line emission. The total continuum bandwidth in D- and E-array configuration observations, including all line-free channels, was 1.53 GHz with mean frequency of 112.7 GHz (corresponding to wavelength of 2.7 mm).  In this paper, we present the $^{12}$CO and continuum maps, and we use $^{13}$CO data to estimate outflow masses (see \S \ref{sec:mass}).  We will present the  $^{13}$CO and C$^{18}$O maps in a future paper discussing the structure of the cloud. 

Integrations were 30 seconds per mosaic position, and 3 minutes on the phase calibrator, with 50 pointings observed in between each phase calibrator observation.  The bright nearby quasar 3C84 was used for phase and gain calibrations, and Uranus was the primary flux calibrator.  Data were reduced using MIRIAD \citep{Sau95}. Anomalous amplitudes and phases were flagged, as were any observations with system temperatures higher than 500 K.  The average system temperature was 310 K.  No shadowed data were used, nor observations with elevation between 85 and 90 degrees.   Flux and gain calibrations were applied, then data from both configurations were Fourier transformed simultaneously.  The maps were inverted using a cell size of 2\arcsec, an image size of 129 pixels for each pointing and a linear mosaic operation, and then deconvolved with the maximum entropy deconvolution \texttt{mosmem}.  The clean images were then restored with a Gaussian beam.  

Since our continuum observations were done simultaneously with the line observations, the maps cover the same area and the observing procedure is the same as described above.  We CLEANed the continuum data using the mosaic-specific task \texttt{mossdi} in MIRIAD, and then restored the clean images with a Gaussian beam.  See Table \ref{tab:carmamaps} for the resulting synthesized beam and rms.  

We also created maps with units of RMS by dividing the intensity map by the sensitivity map (from the MIRIAD task \texttt{mossen}), and correcting for the ratio between the theoretical rms calculated by MIRIAD and the measured rms in the clean image.  This allows us to identify (line) emission features with signal greater than $3\sigma$, given that the sensitivity is not constant \textit{at the edges of} the map since the data were taken with a heterogenous antenna array (see above), and it is particularly useful for identifying significant emission features near the edge of our map.  Maps shown in figures throughout this paper are in units of RMS rather than absolute intensity.

Our maps have best sensitivity (with an rms in the $^{12}$CO map of 0.2 Jy beam$^{-1}$\ per channel) in the regions properly Nyquist sampled and covered by the full array of antennas, approximately the central $\sim6\arcmin\times6\arcmin$ region (i.e. the inner contour shown in Figure \ref{fig:gut08}).  Only a subset of the antennas are sensitive to the outer $\sim1\arcmin$ regions resulting in degraded sensitivity (with an rms in the $^{12}$CO map of 1.1 Jy beam$^{-1}$\ per channel).  Here we report on molecular outflow emission $>3\sigma$ per channel that encompasses a region $\sim 7^\prime \times 7^\prime$ (including part of the map with lower sensitivity), corresponding to an area of $\sim 49$ square arcmin ($\sim 0.23$ pc$^2$). However, it should be kept in mind that there might be undetected outflow emission in the outer $\sim1\arcmin$ of our maps due to the lower sensitivity in this area.

\subsection{Combining Interferometer and Single Dish Line Data} \label{sec:combmethod}

We complement the CARMA observations with $^{12}$CO and $^{13}$CO data in Perseus collected as part of the COordinated Molecular Probe Line Extinction and Thermal Emission (COMPLETE) Survey of Star Forming Regions \citep{Rid06}.  Observations were made with the 14-m Five College Radio Astronomy Observatory (FCRAO) telescope in New Salem, Massachusetts.  For a more detailed description of these observations and data, see \citet{Rid06}.  The FCRAO beamsize at the frequencies of $^{12}$CO and $^{13}$CO  (1-0) is about $46\arcsec$.

\begin{figure}[!ht]
\includegraphics[angle=0,height=0.5\textheight]{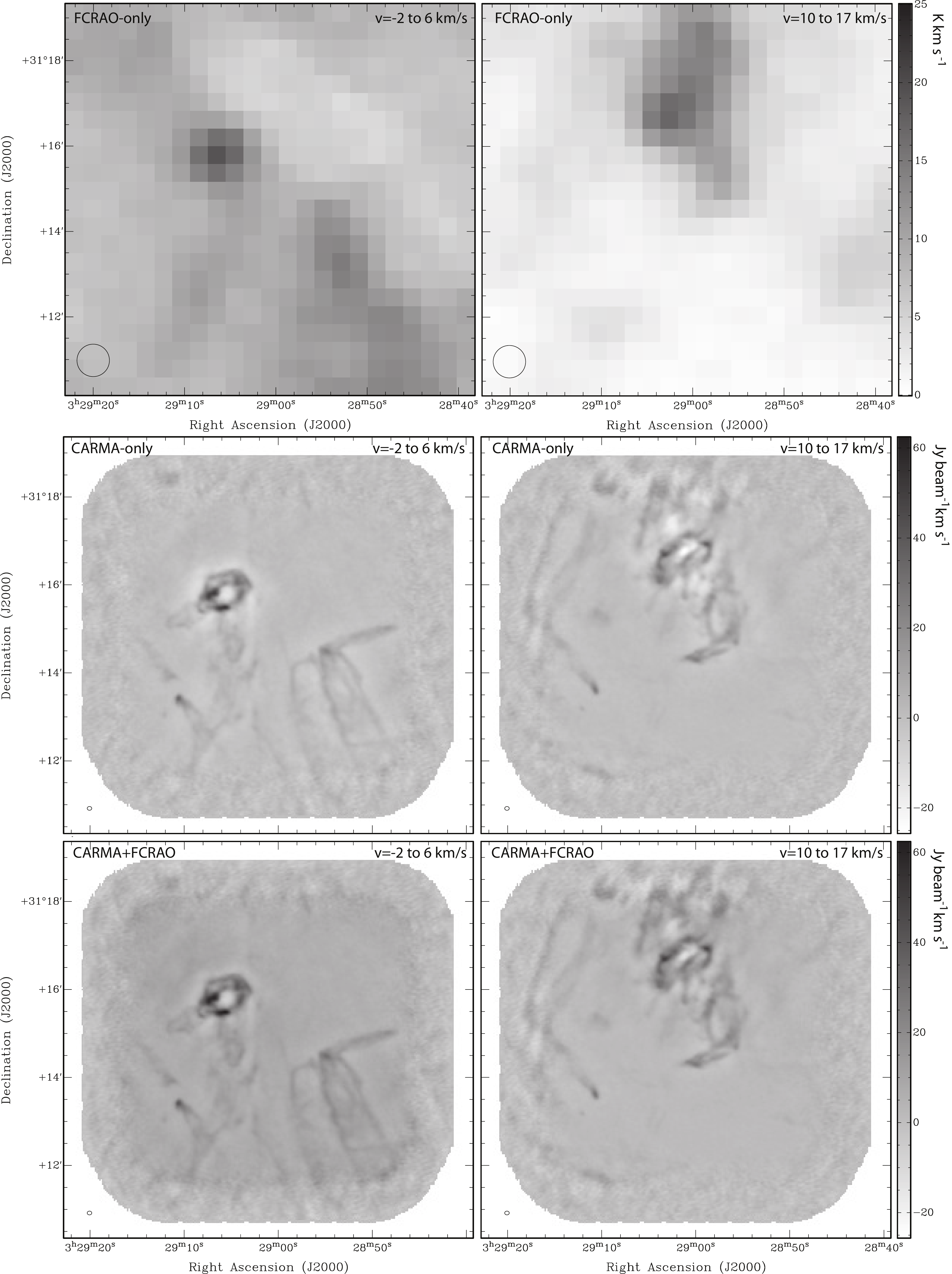}
\caption{Comparison of same region in FCRAO-only (upper), CARMA-only (middle) and combined CARMA+FCRAO (lower) maps.  Left panels show integrated intensity (moment 0) maps of blue-shifted velocity channels between $V_{LSR}=-2$ to $6$ \kms, and right panels show integrated intensity maps of red-shifted velocity channels between $V_{LSR}=10$ to $17$ \kms.  The beam is given in the lower left corner of each panel.  A discussion of the joint deconvolution combination method can be found in \S \ref{sec:combmethod}. }
\label{fig:comparemaps}
\end{figure}

We combine interferometer (CARMA) and single dish (FCRAO) observations in order to recover flux over a range of spatial scales in the region.  The interferometer-only map and the combined map of $^{12}$CO are shown in Figure \ref{fig:comparemaps}. The CARMA map has a resolution of $\sim5^{\prime\prime}$ (or $\sim1000$ AU at a distance of 235 pc), and it is necessary for discerning small-scale structure, particularly important for distinguishing outflows and associating them with their driving sources.  \citet{Eng99} and \citet{Pla00} previously presented BIMA observations of $^{12}$CO in NGC 1333, indicating that interferometer mosaic mapping is a feasible technique for identifying outflow features within the molecular cloud.  Our CARMA observations follow this method, but with more antenna baselines and collecting-area we are now able to achieve higher sensitivity and a more thorough analysis.  The FCRAO map is used to image the larger spatial scales, and recover the flux that is resolved out by the interferometer, including total power which corresponds to zero spatial frequency.  Specifically, this is critical for recovering flux from large-scale outflows, where gas within outflow lobes is typically composed of extended structures.  For a by-eye inspection, we still have the benefit of the interferometer-only map in order to identify certain outflow features, with the interferometer acting as a ``spatial filter'', while the combined interferometer and single dish map allows us to more accurately calculate outflow mass, momentum and energy without filtering out large-scale outflow structure. 

%%%%%%%%%%%%
%fig2
%%%%%%%%%%%%%

The combination of interferometer and single dish CO observations, or the short spacing correction, was carried out according to the non-linear joint deconvolution method of \citet{Sta02}.  This method is particularly useful for recovering flux in mosaics that cover regions with clumpy structure.   The combination was done with MIRIAD.  First we Hanning smoothed the single dish map to a similar velocity resolution as the interferometer map, and then regridded the single dish map using the interferometer map as the template.  Deconvolution was done with \texttt{mosmem}, using the dirty interferometer map and single dish (regridded) map as inputs, along with dirty beams for each.  In order to match the units of the interferometer maps, we converted $T^*_A$ in the single dish maps to units of Jy beam$^{-1}$\, using factors of 21.81 and 21.85 for $^{12}$CO and $^{13}$CO, respectively, and assuming an antenna efficiency of 0.47.  In these units and with the same velocity resolution as the interferometer map we measured an rms in the single dish maps of 5 Jy beam$^{-1}$\ and 2 Jy beam$^{-1}$\ per channel for $^{12}$CO and $^{13}$CO, respectively. For the interferometer maps, we used the beamsize of the CARMA continuum map, which was only slightly larger than that of the $^{12}$CO and $^{13}$CO maps, so that the maps would have equivalent resolution for continuum subtraction and for later calculations (see \S \ref{sec:mass}).

The joint deconvolution was done with the MIRIAD task \texttt{mosmem}.  We use the ``gull'' entropy measure, rather than the cornwell (or maximum emptiness criteria) measure, so as not to force only positive sky-flux solutions and miss negative flux such as absorption below continuum level.  In addition, with \texttt{mosmem} we need to specify several parameters so that the convolution process converges within a reasonable number of iterations, including rms factors for both the interferometer map and single dish map (called ``rmsfac'' in MIRIAD), and the flux calibration factor between the interferometer and single dish map (called ``factor'' in MIRIAD).  The factor ``rmsfac'' is the ratio between the theoretical rms calculated by MIRIAD and the true rms noise, and it is necessary in order for the task to reduce the residuals (i.e. the difference between the dirty image and the model modified by the point spread function) to have the same rms as the theoretical rms multiplied by the ``rmsfac''.  We used ``rmsfac'' of 1.6 and 1.4 for the interferometer maps of $^{12}$CO and $^{13}$CO, respectively.  

The flux calibration factor is the factor by which the single dish data is multiplied in order to convert it to the same scale as the interferometer data.  We determined this factor using the task \texttt{immerge}, choosing uv-data in the range $7.2-12$ meters (which is the range of ``baselines'' that the interferometer and single dish data have in common), and specifying channels where emission structure is comparable in the interferometer and singe dish maps.  This range of baselines corresponds to $3-5$ k$\lambda$. We used a factor of 1.1 for the $^{12}$CO deconvolution, and 0.9 for the $^{13}$CO deconvolution.   

Using these factors, the deconvolution converged within 200 iterations for each channel. The data cubes were then restored with a Gaussian beam.  We note that a joint deconvolution was not performed to combine the CARMA continuum map with single-dish data, since we did not have the appropriate single-dish continuum data available.  Finally, we subtracted the CARMA-only continuum image from the joint deconvolution molecular line maps using \texttt{avmaths} in the image domain. The resulting mean rms sensitivity of the combined CARMA and FCRAO maps of $^{12}$CO  and $^{13}$CO were 0.27 and 0.14 Jy beam$^{-1}$\, respectively, per 0.3 \kms\ channel.

%%%%%%%%%%%%%%%%%%%%%%%%%%%%%%%%%%%%%%%%%%%%
\section{Results}\label{sec:results}

\subsection{Continuum Sources}\label{sec:cont}

\begin{figure}[!ht]
\includegraphics[width=\linewidth,angle=0]{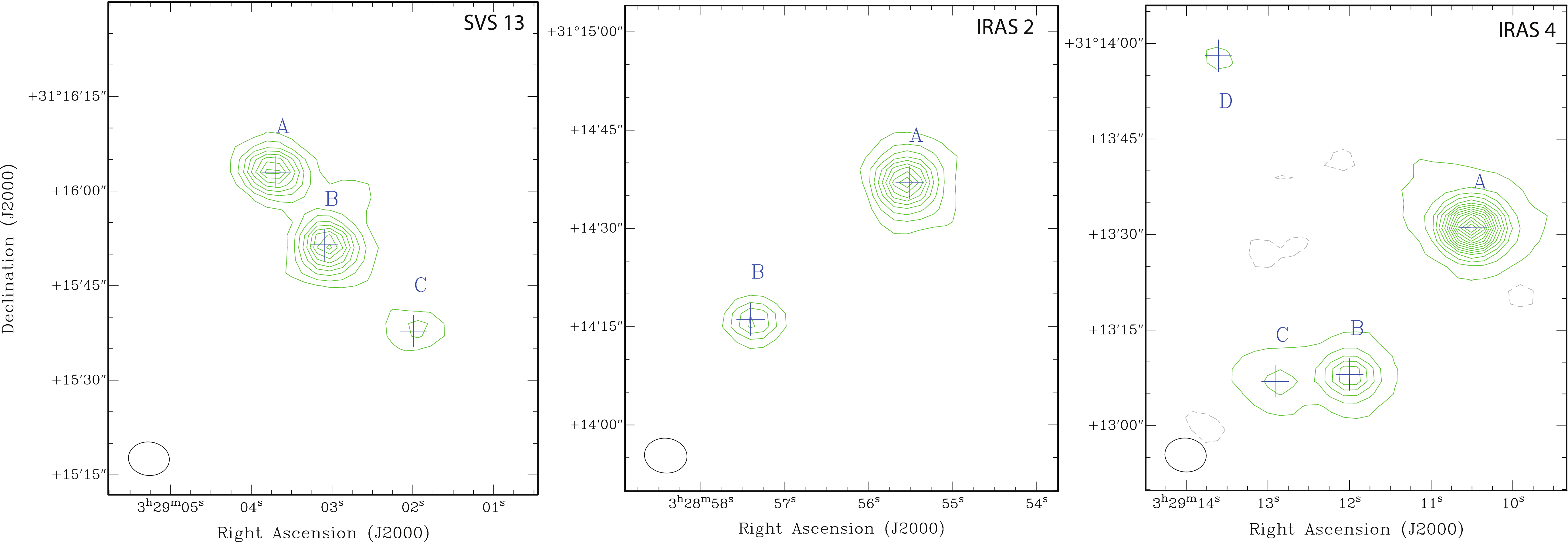}
\caption{Continuum emission at 112.7 GHz in the regions of SVS13, IRAS2 and IRAS4 (regions are outlined in Figure \ref{fig:gut08}).  Green contours begin at 4 $\sigma$ with increments of 4 $\sigma$ for SVS13 and IRAS2, and increments of 20 $\sigma$ for IRAS4 which has the brightest continuum sources in the map.  Gray dashed contours show negative intensity measured at the level of 4 $\sigma$, which only appears in the region of IRAS 4 where the strongest continuum sources are found. The rms of the continuum map is 0.002 Jy beam$^{-1}$.  The synthesized beam is given in lower left corner of each panel.}  
\label{fig:contmap2}
\end{figure}

In our map we detect strong continuum sources in the regions SVS 13, IRAS 2, and IRAS 4 (see Figures \ref{fig:gut08} and \ref{fig:contmap2}), with locations, sizes, peak and total fluxes, and masses of each source given in Table \ref{tab:contsources}. Alternate names and references for these sources in the literature are included in Table \ref{tab:ysos}. Continuum emission likely comes from dust in the circumstellar disks and the protostellar envelopes, and this emission pinpoints the positions of protostars that may, although do not necessarily, drive outflows.   With continuum map rms of 2 mJy beam$^{-1}$\ and beamsize of $6.5\as \times 5.3\as$, we detect 9 continuum sources at greater than $4\sigma$ level, of which we determine that six are driving outflows (some possibly drive multiple outflows).  

%%%%%%%%%%%%%%
%fig3
%%%%%%%%%%%%%%

%%%%%%%%%%%%%%
%tab3
\begin{deluxetable*}{lcccccccccc}
\tabletypesize{\tiny}
\tablecaption{Continuum sources \label{tab:contsources}}
\tablehead{ \colhead{Source} & \colhead{Class}& \multicolumn{2}{c}{Position}&\colhead{Box size \tablenotemark{a}}& \multicolumn{2}{c}{Object Size}&\colhead{PA}  &\colhead{Peak intensity}&\colhead{Total flux} & \colhead{Mass \tablenotemark{b}} \\
\cline{3-4} \cline{6-7} 
\colhead{}&\colhead{}&\colhead{$\alpha$(J2000)}&\colhead{$\delta$(J2000)}  &\colhead{(\arcsec)}& \colhead{Major (\arcsec)}& \colhead{Minor (\arcsec)}& \colhead{(\dg)}& \colhead{(mJy beam$^{-1}$)}&\colhead{(mJy)} &\colhead{(\msun)}}
\startdata
SVS 13A &I & 03:29:03.7 & +31:16:03.1 &20$\times$16& $9\pm1$ & $6\pm1$ & $67\pm12$ & $65\pm6$ & 106 & 0.3\\
SVS 13B & 0 & 03:29:03.1 & +31:15:51.5  &20$\times$18&$9\pm1$ & $8\pm1$ & $39\pm39$ & $68 \pm 7$ & 130 & 0.3\\
SVS 13C & 0 & 03:29:01.9 & +31:15:38.6  &18$\times$10& $11\pm2$ & $7\pm1$ & $85\pm14$ & $16\pm3$ & 35 & 0.09\\
IRAS 2A & 0 & 03:28:55.5 & +31:14:37.0  &22$\times$22&  $8\pm0.5$ &  $8\pm0.5$ & 51 $\pm$117 &  68 $\pm$4 & 129 & 0.3  \\
IRAS 2B & I & 03:28:57.4 & +31:14:15.9  &14$\times$12& $7\pm0.3$ &  $5 \pm0.3$ & $87\pm9$ &  $35\pm2$ & \nodata \tablenotemark{c}  & 0.09 \tablenotemark{c} \\
IRAS 4A & 0 & 03:29:10.5 & +31:13:31.1  &26$\times$26&  $7 \pm0.2$ &  $6\pm0.1$ & $84\pm5$ &  $580 \pm12$ & 773 &  2.0 \\
IRAS 4B & 0 & 03:29:12.0 & +31:13:07.9  &18$\times$16& $7\pm0.4$ & 6 $\pm0.3$ & $88\pm8$ & $226 \pm11$ & 267 & 0.7 \\
IRAS 4C & 0 & 03:29:12.9 & +31:13:06.9  &14$\times$14& $8\pm1$ & $6\pm0.4$ & $82\pm9$ & $78\pm5$ & 110 & 0.3 \\
IRAS 4D & 0 & 03:29:13.6 & +31:13:57.9  &10$\times$10&  $6\pm2$ & $4\pm1$ & $65\pm20$ & $20\pm5$ & \nodata \tablenotemark{c}  & 0.05\tablenotemark{c} \\  
\enddata
\tablenotetext{a}{Box size refers to the region defined around the continuum source, used as input for Miriad task imfit and described in \S \ref{sec:cont}.}
\tablenotetext{b}{Mass calculation is described in \S \ref{sec:cont} and follows the method of \citet{Loo00}.}
\tablenotetext{c}{Miriad task imfit failed to fit a gaussian to these sources.  For these sources, we find the mass (lower limit) based on peak intensity. }
\end{deluxetable*}
%%%%%%%%%%%%%%

With this sensitivity, we are able to detect at greater than a $4\sigma$ level envelope masses greater than 0.024 \msun\ for sources within the region of best sensitivity, according to the following mass relation \citep[see][]{Sch10}:
\begin{equation}
M=\frac{d^2 S_\nu}{B_\nu(T_D) \kappa_\nu},
\end{equation}
where $d$ is distance to the source, $S_\nu$ is the 2.7 mm continuum flux, and $B_\nu(T_D)$ is the Planck function.  We assume a dust temperature of $T_D=30$ K \citep{Che13}, and a dust opacity of $\kappa_\nu=0.1(\nu/1200\textrm{ GHz})$\ cm$^2$g$^{-1}$, which corresponds to $\kappa_\nu=0.009$\ cm$^2$g$^{-1}$ at $\lambda=2.7$\ mm, following \citet{Loo00}.  We note that dust temperature and opacity in particular contribute to the uncertainty of the masses calculated from continuum. Nonetheless, our mass results reported in Table \ref{tab:contsources} are consistent within a factor of a few with those of other works based on interferometer observations \citep{Loo00,Cho01,Che09,Che13}, correcting for assumed distance, and other parameters described above. 

Following \citet{Loo00}, we acknowledge that the best estimates of masses require detailed modeling, which we do not perform here.  We must consider that the integrated flux we measure is likely correlated with beamsize and box size, as noted by \citet{Lad91}.  \citet{Loo00} present this effect for several of the sources that we present here, finding that for these sources the integrated flux differs by a factor of about 2-3 depending on the box size they choose (see their Table 2). For clarity, we provide in Table \ref{tab:contsources} the size of the box we defined around each continuum source.  We chose box sizes to include all significant emission surrounding the point sources, without including nearby sources.  As a simple test, we calculated integrated flux for several of the relatively more isolated sources (IRAS 2A, IRAS 2B and IRAS 4A) using a larger box size, and we find very little effect on the results in these cases.  For other more closely-spaced objects (i.e. the SVS 13 region, see \S \ref{sec:svs13_cont}), some of which may perhaps share a common envelope, expanding the box size clearly has an effect if extended emission is more significant.    

It is likely that some sources below the detection limit of our continuum observations also drive outflows, such as the candidate outflows presented in \S \ref{sec:cand_out}.  In particular, we do not detect significant continuum emission at the locations of the sources SK 1 and SK 14 \citep{San01}, which are compact 450 $\mu$m and 850 $\mu$m sources with mass estimates of 0.07 \msun\ and 0.01 \msun, respectively, according to \citet{San01}, but which drive candidate outflows (see \S \ref{sec:cand_out}).   We do not detect the continuum source SK 14 because it has a mass which is below the detection limit of our map.  We suspect that we also do not detect the continuum source SK 1 because it lies at the southern edge of our map, where the 4$\sigma$ sensitivity is 0.054 mJy beam$^{-1}$, which translates to a mass detection limit of about 0.06 \msun (using the same assumptions as above), and the previously reported mass of this object is very close to this limit.  In Figure \ref{fig:contmap2} we feature regions where we detected significant continuum emission in our map.  In the following sub-sections, we describe characteristics of the individual continuum sources, as a preface to the discussion of outflow characteristics in \S \ref{sec:molecules}-\ref{sec:cand_out}.  

Further, we note that three starless cores are located in the mapped region (Bolo 44, SK 16 and HRF 59; see Table 2, as well as Hatchell et al. 2007a), but we do not detect significant continuum emission at any of their locations.  SK 16 ($M_{env}=8.4$ \msun), located southeast of SVS 13, may have comparable mass to the lowest-mass cores in our map.  However, \citet{Sch12} suggest that starless cores may not be detected by interferometer observations if their density distribution is sufficiently flat and their continuum emission smooth, and this is very likely the case for SK 16.  The other two starless cores (Bolo 44 and HRF 59) lie in the very low sensitivity (1$\sigma$ rms  of 12 mJy beam$^{-1}$) region of our map and we do not detect them as their intensity is lower than our 4$\sigma$ sensitivity limits.  

\subsubsection{SVS 13 Region}\label{sec:svs13_cont}

Within the region shown in the left panel of Figure \ref{fig:contmap2} lies the near-IR source discovered by \citet{Str76} known as SVS 13, and later observed with IRAS by \citet{Jen87} (i.e., NGC 1333 - IRAS 3), which is responsible for powering HH 7-11 and the corresponding molecular outflow (see \S \ref{sec:svs13_out}).  The millimeter continuum observations by \citet{Chi97} showed three distinct sources in the region -- which they name MMS1, MMS2 and MMS3 and are more commonly known as SVS 13A, SVS 13B and SVS 13C.  Of these, SVS 13A corresponds to the optical/NIR source SVS 13.  Sensitive, high-angular resolution VLA observations at 3.6 cm and 6 cm \citep{Rod97,Rod99} showed that the millimeter sources correspond to radio sources (VLA 4, 17 and 2), and they detected a fourth source (VLA 3), southwest of SVS 13A.  In addition, SVS 13C is associated with an H$_2$O maser source known as H$_2$O(B) \citep{Has80,Hir08}.  \citet{Loo00} observed the 2.7 mm continuum emission using BIMA and detected four sources in this region, including SVS 13B, SVS 13C, and resolving SVS 13A into the two components SVS 13A1 and SVS 13A2 (at the position of VLA 3) separated by 6\as.  \citet{Che09} suggest that VLA 3 (their naming convention for the source SVS 13A2) in fact forms a protobinary system with SVS 13B based on the velocity field deduced from IRAM PdBI observations, even though VLA 3 appears to be nearer to SVS 13A than SVS 13B as seen projected on the plane of the sky.  Encompassing these sources is a large-scale common envelope which was detected with single-dish mm/sub-mm observations \citep{Chi97,Cha00}, but resolved out with interferometer observations. 

In this region, we detect the three distinct continuum point sources SVS 13A, SVS 13B and SVS 13C shown in Figure \ref{fig:contmap2}. We cannot resolve VLA 3, but we suggest that the elongated emission near SVS 13A and SVS 13B is likely related to this source. Based on our continuum map, sources SVS 13B and SVS 13C lie southwest of SVS 13A at distances $15 \arcsec$ (0.02 pc) and $36 \arcsec$ (0.04 pc), respectively, along a line with position angle 230\dg (see Figure \ref{fig:contmap2}).  Most studies classify SVS 13A as a Class I source \citep[e.g.][]{Hat07a,Eva09}.  Based on its bolometric temperature ($\sim114$ K) and ratio of sub-mm to bolometric luminosity ($\sim0.8$\%) \citet{Che09} also suggest SVS 13A is Class I, but they note that its association with a cm source and its high-velocity outflow are characteristic of a younger, Class 0 object. Hence, they speculate that SVS 13 A could be a transition Class 0/I source (i.e., a very young Class I source).  Both SVS 13B and SVS 13C have been shown to be Class 0 sources based on their SEDs and mm/sub-mm envelope properties \citep{Cha00,Che09}. 

\subsubsection{IRAS 2 Region}\label{sec:IRAS2_cont}

IRAS 2 was discovered as a protostellar candidate based on IRAS observations by \citet{Jen87}, and two counterpart sources have been observed at radio, sub-mm and mm wavelengths \citep{Rod99,San94,Loo00}.  The northwest source is known as IRAS 2A, and the southeast source is IRAS 2B (see Figure \ref{fig:contmap2}).  The near-IR source SVS 19 is very nearly coincident with IRAS 2B, however \citet{Rod99} suggest SVS 19 corresponds to the bright G2 IV star in the foreground, BD +30\dg547 (a.k.a., ASR 130).  In addition to IRAS 2A and IRAS 2B, \citet{San01} detected the source IRAS 2C northwest of IRAS 2A.

In our map we detect two continuum sources separated by $35 \arcsec$ (0.04 pc) in the IRAS 2 region, shown in Figure \ref{fig:contmap2} and previously detected by \citet{Loo00}.  IRAS 2A is a Class 0 protostar, while IRAS 2B is likely a young Class I source, based on its bolometric temperature ($T_{bol}=100$) and spectral index ($\alpha=1.48$) \citep{Eva09}.  However, \citet{San01} called IRAS 2B a possible Class 0 source, and \citet{Gut08} classify it as deeply embedded.  Since IRAS 2A drives nearly perpendicular outflows, described in more detail in \S \ref{sec:iras2_out}, it is likely that this is a binary source \citep[][and references therein]{Che13}, although we detect only one source. \citet{Che13} report a separation between the components of IRAS 2A of 1\farcs5 and a mass ratio of 0.05 according to SMA 850$\mu$m dust continuum results.  

\subsubsection{IRAS 4 Region}\label{sec:IRAS4_cont}

IRAS 4 was also observed in the IR by \citet{Jen87}, with the three radio counterparts VLA 25, VLA 28 and VLA 29 observed by \citet{Rod99}.  VLA 25 and VLA 28 correspond to IRAS 4A and IRAS 4B in Table \ref{tab:ysos}. \citet{San01} resolved VLA 28 into two components, which correspond to the sources commonly known as IRAS 4B and IRAS 4C (these are IRAS 4BW and IRAS 4BE in Sandell \& Knee, 2001).  VLA 29 corresponds to the source we call IRAS 4D. \footnote{This naming convention is chosen to abide by the IRAS 4A, 4B, 4C sequence used by \citet{Loo00} and \citet{Che13}, although it should be noted that the source we call IRAS 4D is called IRAS 4C in several other works \citep{Lef98,San91}, and the sources we call IRAS 4B and 4C are called, e.g., IRAS 4BW/E \citep{San01} or IRAS4BI/II \citep{Cho01}. }

Figure \ref{fig:contmap2} shows the four continuum sources, IRAS 4A, IRAS 4B, IRAS 4C and IRAS 4D, of which IRAS 4A and IRAS 4B are the strongest continuum sources detected in our map.  IRAS 4B is located $33\arcsec$ (0.04 pc) southeast of IRAS 4A, and IRAS 4C is located $14\arcsec$ (0.02 pc) east of IRAS 4B.  IRAS 4D is a weaker (with a $7\sigma$ peak) source that lies $54\arcsec$ (0.06 pc) northeast of IRAS 4A.  The four IRAS 4 sources have been classified as Class 0 \citep{San01,Hat07a,Hat09,Eva09}.  High angular-resolution continuum observations have shown that IRAS 4A is a binary source \citep{Lay95,Loo00,Che13} whose components have a separation of 1\farcs4 and mass ratio of 0.5. Given the angular resolution of our CARMA observations we cannot resolve the components that make up IRAS 4A in our map.  Previous works have also shown that IRAS 4B and IRAS 4C are components of a protostellar-multiple system \citep{Lay95,Loo00,Che13}.   

\subsection{Molecular Outflow Emission} \label{sec:molecules}

Here we present CO (1-0) outflows associated with SVS 13, IRAS 2, and IRAS 4, followed by several outflow candidates in \S \ref{sec:cand_out}.  We also compare with previous observations of molecular outflows across this region, including single-dish maps of CO (1-0) by \citet{Kne00} and CO (3-2) by \citet{Kne00} and \citet{Cur10a,Cur10b}, \footnote{The maps of \citet{Kne00} are $\sim63$ arcmin$^2$, and the survey was later expanded upon by \citet{Hat07b} and \citet{Hat09}. \citet{Cur10a,Cur10b} mapped a region 612 arcmin$^2$ in NGC 1333.} as well as interferometry studies of individual outflows within the region, mentioned more specifically in the following sub-sections.

\begin{figure*}[!ht]
\includegraphics[width=\linewidth]{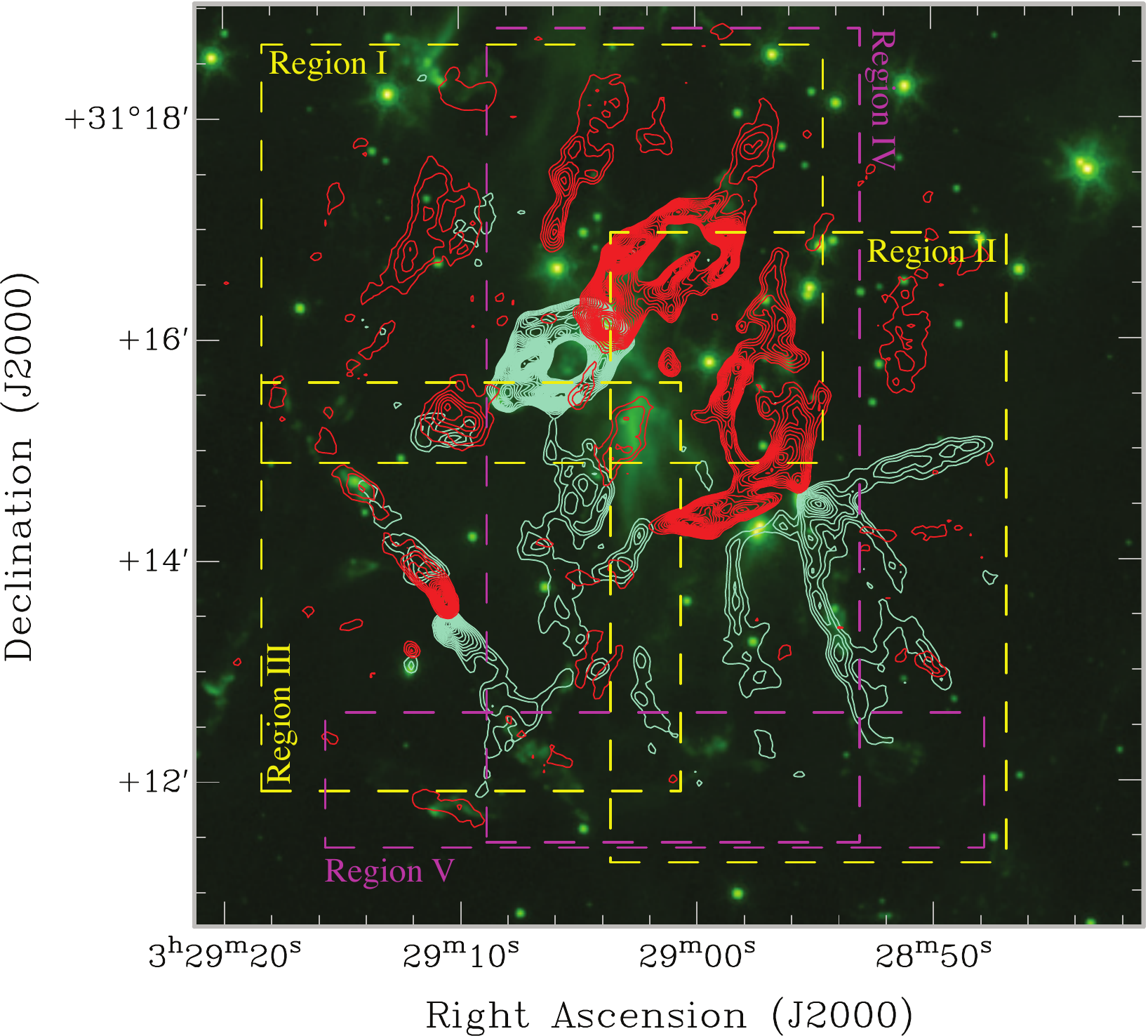}
\caption{$^{12}$CO emission in the mapped region of NGC 1333 (contours), overlaid on a map of IRAC $4.5\mu$m emission \citep{Gut08}.  Contours show integrated intensity of blue- and red-shifted emission in channels with velocity $V_{LSR}=-2.3$ to $6.2$ \kms (blue) and $V_{LSR}=10.0$ to $17.3$ \kms (red).  Contours begin with 2$\sigma$ and increment by $2\sigma$, where $\sigma$ is the rms of the respective moment map. We detect outflow emission greater than $3\sigma$ spanning a region of approximately $7\arcmin\times7\arcmin$ in size (see \S \ref{sec:carmaobs}).  Dashed boxes indicate the regions we show with more detail in Figures \ref{fig:svs13}-\ref{fig:candsk1} and are described in more detail in \S \ref{sec:molecules}-\ref{sec:cand_out}.  In the following figures, we show integrated intensity maps for low-, mid-, and high-velocity CO emission in Regions I-III (yellow dashed boxes), and for Regions IV-V (magenta dashed boxes) we show just the blue- and red-shifted velocity channels where most outflow emission is found.}
\label{fig:wholemap}
\end{figure*}

Our map of $^{12}$CO outflow emission is shown in Figure \ref{fig:wholemap}, with outflows detected within a region spanning $\sim 7^\prime \times 7^\prime$.   These cool outflows are traced well with $^{12}$CO, which probes the entrained molecular gas and allows us to determine physical outflow properties.  In these regions, the $^{13}$CO is less optically thick than the $^{12}$CO, and it was used to make a reliable estimate of the outflow mass (see \S \ref{sec:mass}).  Table \ref{tab:areavol} gives physical traits of the outflows, including position angles and sizes. In the following sub-sections we describe their distinct morphologies in more detail.  Table \ref{tab:sources} gives quantitative characteristics including mass, momentum and energy for the outflows in the region, as well as the outflows' driving sources.  The method to calculate mass, momentum and energy is described in \S \ref{sec:mass}.  

%%%%%%%%%%%%%%%%
%fig 4
%%%%%%%%%%%%%%%%

%%%%%%%%%%%%%%%%
%tab4
\begin{deluxetable*}{lccccccc}
%\rotate
%\tablewidth{8in}
%\setlength{\tabcolsep}{0.01in} 
%\tabletypesize{\tiny}
\tablecaption{Outflow Lobe Physical Traits \label{tab:areavol}}
\tablehead{
\colhead{Name   }&\colhead{PA    }&\colhead{$a$     }&\colhead{$b$     }&\colhead{coll. \tablenotemark{a} }&\colhead{ $\theta$ \tablenotemark{b}}&\colhead{A\tablenotemark{c} }&\colhead{V\tablenotemark{c}}\\
 &\colhead{(deg) }&\colhead{(\as) }&\colhead{(\as) }&\colhead{ ($a/b$) }&\colhead{(deg)   }&\colhead{($10^3$ arcsec$^2$)    }&\colhead{($10^{-4}$pc$^3$)} }
\startdata
Blue Lobe	&&&&&&&\\
\hline
SVS 13A\tablenotemark{d}&  120&  122&   82&   0.7&   96&    7.88&   6.39\\
SVS 13C&    8&  212&   86&   0.4&   69&   14.41&  12.28\\
IRAS 2A west-east&  103&  108&   21&   0.2&   30&    1.74&   0.35\\
IRAS 2A south-north&   24&  188&   70&   0.4&   56&   8.71\tablenotemark{e}&   6.56\tablenotemark{e}\\
IRAS 2B \tablenotemark{f}&   24&  181&   92&   0.5&   60&   12.99&   11.73 \\
IRAS 4A\tablenotemark{g}&   35&  104&   24&   0.2&   100&    2.00&   0.48\\
IRAS 4B&  180&   10&   10&   1.0&   90&    0.08&   0.01 \\
SK 14&  145&   71&   39&   0.6&   93&    2.16&   0.83 \\
SK 1&   98&  165&   32&   0.2&   \nodata&    1.63\tablenotemark{h}&   0.58\tablenotemark{h}\\
C1&\nodata&\nodata&\nodata&\nodata&\nodata&\nodata&\nodata\\
C2&148&  122&   34&   0.3&   \nodata&    3.25&   1.09 \\
C3&\nodata&\nodata&\nodata&\nodata&\nodata&\nodata&\nodata\\
C4&\nodata&\nodata&\nodata&\nodata&\nodata&\nodata&\nodata\\
\hline
Red Lobe	&&&&&&&\\
\hline
SVS 13A\tablenotemark{d}&  140&  163&   57&   0.4&   76&    7.35&   4.16\\
SVS 13C&    8&  180&   86&   0.5&   60&   12.21&  10.41\\
IRAS 2A west-east&  105&   84&   22&   0.3&   32&    1.48&   0.33\\
IRAS 2A south-north&    4&  133&   90&   0.7&   93&    9.39&   8.33\\
IRAS 2B \tablenotemark{f}&    \nodata&  \nodata&   \nodata&   \nodata&   \nodata&    \nodata&   \nodata\\
IRAS 4A\tablenotemark{g}&   35&  125&   29&   0.2&   82&    2.87&   0.83\\
IRAS 4B&  180&   11&   11&   1.0&   90&    0.09&   0.01\\
SK 14&  145&   58&   35&   0.6&   55&    1.59&   0.55\\
SK 1&   95&  165&   32&   0.2&   \nodata&    2.04\tablenotemark{h}&   0.72\tablenotemark{h}\\
C1&157&  180&   36&   0.2&   \nodata&    5.09&   1.81\\
C2&\nodata&\nodata&\nodata&\nodata&\nodata&\nodata&\nodata\\
C3&  158&  202&   65&   0.3&   \nodata&   10.26&   6.56\\
C4&  160&  144&   18&   0.1&   \nodata&    2.04&   0.36\\
\hline
Blue and Red Average or Sum	&&&&&&&\\
\hline
SVS 13A\tablenotemark{d}&  130&  143&   70&   0.6&   86&   15.23&  10.55\\
SVS 13C&    8&  196&   86&   0.5&   65&   26.63&  22.68\\
IRAS 2A west-east&  104&   96&   22&   0.3&   31&    3.22&   0.68\\
IRAS 2A south-north&   14&  161&   80&   0.6&   75&   18.10 &  14.89\\
IRAS 2B \tablenotemark{f}&   \nodata&  \nodata&   \nodata&   \nodata&   \nodata&   \nodata&  \nodata\\
IRAS 4A\tablenotemark{g}&   35&  115&   27&   0.2&   91&    4.87&   1.31\\
IRAS 4B&  180&   11&   11&   1.0&   90&    0.17&   0.018\\
SK 14& 145&   65&   37&   0.6&   74&    3.75&   1.38\\
SK 1&   97&  165&   32&   0.2&   \nodata&    3.67&   1.3\\
C1&\nodata&\nodata&\nodata&\nodata&\nodata&\nodata&\nodata\\
C2&\nodata&\nodata&\nodata&\nodata&\nodata&\nodata&\nodata  \\
C3&\nodata&\nodata&\nodata&\nodata&\nodata&\nodata&\nodata\\
C4&\nodata&\nodata&\nodata&\nodata&\nodata&\nodata&\nodata\\
\enddata
\tablenotetext{a}{Collimation factor $b/a$, where $b$ is width and $a$ is length of outflow lobe.}
\tablenotetext{b}{Opening angle $\theta$ fit by eye.  Fit was only made for outflows with identified sources, and where emission was detected near the identified source.}
\tablenotetext{c}{Area (A) and volume (V) of ellipse fit to outflow lobe, assuming idealized, symmetric outflow geometry, as in \citet{Off11}. For volume, we assume that depth equals width of outflow lobe ($c=b$). }
\tablenotetext{d}{``Blue'' lobe of SVS 13A is in southeast, and includes red channels $9.08-11.30$ \kms, in addition to blue channels.}
\tablenotetext{e}{ Area and volume of IRAS 2A-south reported after subtracting area and volume of SK 1-west, where the two lobes overlap.  See \S \ref{sec:iras2_out} for more discussion. }
\tablenotetext{f}{ We did not disentangle emission from nearby contaminating outflows to estimate mass for IRAS 2B (see \S \ref{sec:iras2_out}). }
\tablenotetext{g}{ ``Red'' lobe of IRAS 4A is in northeast, and includes blue channels $2.73-5.91$ \kms, in addition to red channels.  }
\tablenotetext{h}{Area and volume are calculated only for the emission features marked in Figure \ref{fig:candsk1}.  PA, $a$ and $b$ for SK 1 are based on the assumption that these features extend from the source SK 1 to form the outflow lobes. }
\end{deluxetable*}
%%%%%%%%%%%%%%%%

%tab5

\begin{deluxetable*}{lcccccccccccccccc}
%\rotate
\tabletypesize{\footnotesize}
%\tablewidth{0pt}
\setlength{\tabcolsep}{0.01in} 
\tablecaption{Outflows and their sources \label{tab:sources}}
\tablehead{ &\multicolumn{4}{c}{Driving Source\tablenotemark{a}}&&\multicolumn{3}{c}{Blue Lobe} &&\multicolumn{3}{c}{Red Lobe} &&\multicolumn{3}{c}{Sum} \\
\cline{2-5} \cline{7-9} \cline{11-13} \cline{15-17}
\colhead{Name} &\colhead{YSO}&\colhead{$\alpha$}&\colhead{$T_{B}$ }&\colhead{$L_{B}$}&&  \colhead{Mass}&  \colhead{Momentum}&    \colhead{Energy}& \colhead{}&  \colhead{$M$}&  \colhead{$P$}&    \colhead{$E$}&  \colhead{}&  \colhead{$M$}& \colhead{$P$}&    \colhead{$E$}\\
\colhead{} &\colhead{Class} &\colhead{} &\colhead{(K)} &\colhead{($L_\odot$)} &&   \colhead{(\msun)}&  \colhead{(\msun\kms)}&    \colhead{($10^{43}$ erg)}&\colhead{}&   \colhead{}&  \colhead{}&    \colhead{}&\colhead{}&    \colhead{}&  \colhead{}&    \colhead{} } 
\startdata
SVS 13A\tablenotemark{b} 		 & I & 1.02  &250 & 59 &&  1.0&  2.7&    9.8&&  0.7&  2.2&    7.8&&  1.7&  4.6&   17.6\\
SVS 13C 		 & 0 &\nodata&36\tablenotemark{c}	& 4.9\tablenotemark{c} && 0.8&  2.5&    9.7&&  1.1&  3.3&   12.1&&  1.8&  5.0&   21.8\\
IRAS 2A-WE	 & 0 &2.71  & 57 & 76  &&  0.1&  0.3&    1.4&&  0.1&  0.2&    0.8&&  0.2&  0.6&    2.1\\
IRAS 2A-SN\tablenotemark{d}		& 0 &2.71  & 57 & 76  &&  0.5	&  1.8	&    8.1	&&  0.4&  1.3&    4.7&&  0.9&  3.7&   12.8\\
IRAS 4A\tablenotemark{e}		& 0 &2.55  & 43   & 5.8  &&  0.2&  0.6&    2.3&&  0.1&  0.4&    1.4&&  0.3&  1.7&    3.7\\
IRAS 4B 		& 0 &0.87  & 55 &1.1 &&  0.01&  0.04&    0.2&&  0.01&  0.03&    0.1&&  0.02&  0.1&    0.3\\
SK 14 		& 0 &1.41  & 59 & 0.2 &&  0.1&  0.4&    1.3&&  0.1&  0.1&    0.5&&  0.2&  0.7&    1.8\\
SK 1 		& 0 &2.07  & 32 & 0.7 &&  0.1&  0.5&    2.5&&  0.1&  0.2&    0.6&&  0.2&  1.0&    3.1\\
C1 			& \nodata &\nodata &\nodata & \nodata && \nodata & \nodata & \nodata &&  0.1&  0.2&    0.7&&0.1&  0.2&    0.7\\
C2 			& \nodata &\nodata &\nodata & \nodata&&  0.2&  0.7&    3.3&& \nodata & \nodata & \nodata && 0.2&  0.7&    3.3\\
C3 			& \nodata &\nodata &\nodata & \nodata&&   \nodata & \nodata & \nodata &&  0.3&  0.8&    2.8&& 0.3&  0.8&    2.8\\
C4 			& \nodata &\nodata &\nodata & \nodata&&   \nodata & \nodata & \nodata &&   0.2&   0.5&    1.6&&  0.2&   0.5&    1.6\\
\hline\\
Sum	&	&		&		&		&&  3.0&  9.5&   38.4&&  3.0&  9.3&   33.3&&  6.0& 18.8&   71.7\\
\enddata

\tablenotetext{a}{ YSO class, bolometric temperature ($T_{bol}$), bolometric luminosity ($L_{bol}$) and slope of SED between 2 and 24 $\mu$m ($\alpha$) pertain to identified outflow-driving sources from Evans et al. (2009), except where marked otherwise.}
\tablenotetext{b}{``Blue'' lobe of SVS 13A includes red channels $9.08-11.30$ \kms, in addition to blue channels.}
\tablenotetext{c}{$T_{bol}$ and $L_{bol}$ from Chen et al. (2009)}
\tablenotetext{d}{Mass, momentum and energy of IRAS 2A-south reported after subtracting the values for SK1-west, where the two lobes overlap.  We also suspect that IRAS 2A-south may be contaminated by IRAS 2B-south, and IRAS 2A-north may be contaminated by IRAS 2B-north, since we can't disentangle the IRAS 2B outflow.  However, the majority of emission appears to be associated with IRAS 2A.  See \S \ref{sec:iras2_out} for more discussion. }
\tablenotetext{e}{ ``Red'' lobe of IRAS 4A includes blue channels $2.73-5.91$ \kms, in addition to red channels.  }
\end{deluxetable*}

%%%%%%%%%%%%%%%%
Our full velocity coverage was $v_{LSR}=-2.3$ to $17.3$ \kms, and we group channels as: blue high-velocity $v_{LSR}=-2.3$ to $-0.4$ \kms, blue medium-velocity $v_{LSR}=-0.1$ to $3.1$, blue low-velocity $v_{LSR}=3.4$ to $5.9$ \kms, red low-velocity $v_{LSR}=9.7$ to $11.0$ \kms, red mid-velocity $v_{LSR}=11.3$ to $14.5$ \kms \ and red high-velocity $v_{LSR}=14.8$ to $17.3$ \kms, chosen to show the predominant outflow features.  Here we assume a cloud velocity of $v_{cloud}=8$ \kms, in agreement with the velocity reported by \citet{Cur10b}. Throughout, we will use the notation of $v_{LSR}$ to indicate channel velocity (relative to local standard of rest), cloud velocity $v_{cloud}$, and outflow velocity $v_{out}=v_{LSR}-v_{cloud}$ (along the line of sight).

%%%%%%%%%%%%%%%%%%%%%%%%%%%%%%%%%%%%%%%%%%%
\subsubsection{SVS 13A Outflow}\label{sec:svs13_out}

%%%%%%%%%%%%%%%%
%fig5
\begin{figure*}[!ht]
\includegraphics[width=\linewidth]{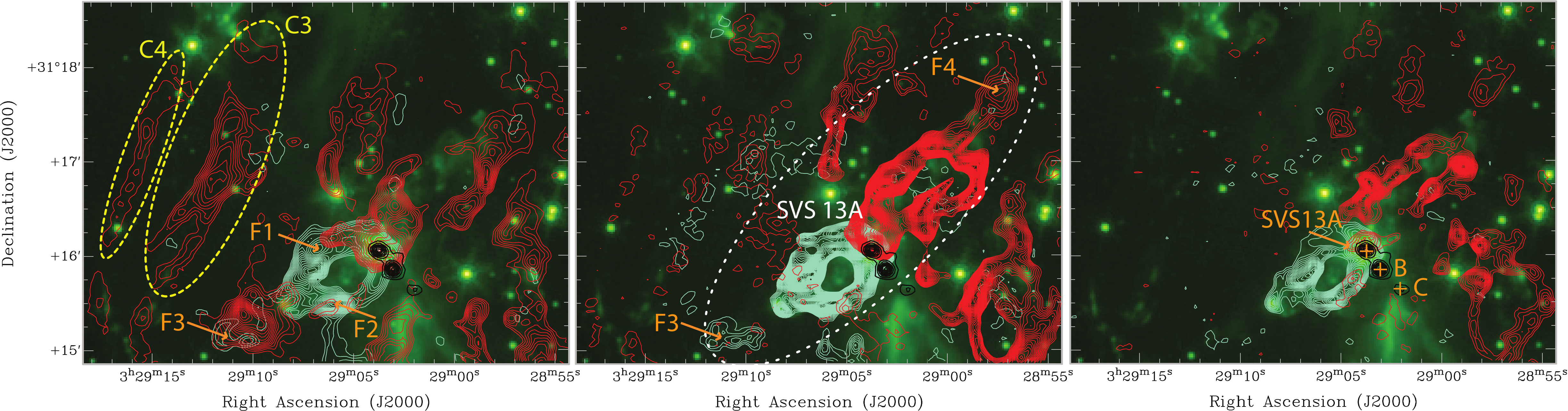}
\caption{Region I  low-, mid-, and high-velocity outflows (left, middle, and right panels, respectively).  Blue- and red-shifted CO outflows are shown with blue and red contours, respectively, continuum emission is shown with black contours, and background is IRAC 4.5$\mu$m emission \citep{Gut08}.  Continuum sources SVS 13A, SVS 13B and SVS 13C are marked with plus signs and labeled only in the right panel. Outflows in this region are approximately outlined with ellipses, to guide the eye.  The SVS 13 outflow lobes (white dotted ellipse) have ``donut'' morphologies and are driven by SVS 13A. F1 and F2 mark the locations of the strongest clumpy emission features in our map, and F3 and F4 mark two distinct ``blobs'' of emission at the tips of both lobes.  Located northeast of SVS 13 are the candidate outflows C3 and C4 (yellow dashed ellipses), seen predominantly in low-velocity red-shifted emission. We group channels as described at the beginning of \S \ref{sec:molecules}.  Contours begin with $3 \sigma$ and increment by $3 \sigma$ for each CO integrated intensity map, and begin with $4 \sigma$ and increment by $4 \sigma$ for continuum.  }
\label{fig:svs13}
\end{figure*}
%%%%%%%%%%%%%%%%

The most prevalent outflow in our map is SVS 13A, shown as Region I of Figure \ref{fig:wholemap} and featured in Figure \ref{fig:svs13}.  Wide-angle blue- and red-shifted outflow lobes are oriented SE-NW and intersect at the location of the young Class I source SVS 13A, which was also suggested by \citet{Kne00} to be the driving source of this outflow.  In most channels, we see the morphology of each lobe as a ``donut''-shape, with strong emission around the edges, and a cavity without emission at the center of each lobe.  Near the driving source SVS 13A, blue- and red-shifted emission from the SE and NW lobes coincide along the line of sight, in agreement with a moderate inclination angle \citep[e.g. 40$^{\circ}$ with respect to the line of sight,][]{Dav11}.  We note that some of the lack of emission detected in the center of the outflow shells is likely due to destructively interfering negative side-lobes (i.e., not all flux is fully recovered in our maps).  The outflow emission from this source has a broad morphology, with blue- and red-shifted lobes having opening angles of 96\dg \ and 76\dg \ and collimation factors of 0.7 and 0.4, respectively.  This wide-angle morphology is consistent with other outflows powered by Class I protostars \citep[e.g.][]{Arc06}.  

Within our spectral coverage, the morphology of the outflow appears to be velocity dependent, with a more clumpy structure seen in the high-velocity channels, and a smoother shell-like structure seen in low-velocity channels.  The low-velocity outflow morphology is reminiscent of L1551, for which \citet{Mor87} suggested that the low-velocity outflow gas is entrained as an expanding shell at the outflow cavity edges, while higher-velocity outflow gas is more concentrated along the outflow axis.  In the SVS 13A outflow, \citet{Bac00} previously identified high-velocity ``molecular bullets" of CO(2-1) emission at velocities greater than about 75 \kms\ from the cloud velocity, and as expected we detect emission over the full extent of our velocity coverage (approximately $v_{cloud} \pm 10$ \kms).  Our limited bandwidth does not allow us to detect the high-velocity ($|v_{out}| >10$ \kms) emission for this outflow, and therefore our mass, momentum and kinetic energy estimates for the SVS 13A outflow should be considered as lower limits. We see strong ``clumpy'' emission features especially in the southeast lobe, at the locations marked F1, F2 and F3 in Figure \ref{fig:svs13}, where F1 and F2 show strong (up to $\sim60\sigma$) blue-shifted emission at $V_{LSR}=3.1-4.6$ \kms, and F3 shows strong ($>40 \sigma$) red-shifted emission at $V_{LSR}=10.3-10.7$ \kms, as well as less-intense blue-shifted emission.  These are the strongest, most concentrated CO emission clumps in our map.  

Comparing our map with previous millimeter interferometric observations of this source \citep[e.g.][]{Bac00}, it can be seen that fully mapping the outflow area and velocity extent is necessary to reveal the complex morphology of this source (and of other wide-angle outflows). With our spatial coverage we recover the full shell-like structure of both outflow lobes, which were previously only partially seen by \citet{Bac00}, and we clearly see that these lobes intersect at the position of the source SVS 13A.

Spitzer IRAC Band 2 observations of this region \citep{Gut08} reveal strong knotty 4.5$\mu$m emission, which is thought to arise from shock-heated molecular hydrogen \citep[e.g.][]{Neu08}.  In the SVS 13A outflow, the bright IRAC 2 emission likely indicates the region where the protostellar wind collides with ambient cloud material, particularly near the SE and NW apexes of the outflow lobes.  Apart from the main contiguous SE-NW CO outflow lobes, there are two distinct ``blobs'' of $^{12}$CO emission to the southeast of the southern lobe (clump F3, previously mentioned) and northwest of the northern lobe (clump F4).  The clumpy morphology of the outflow and the fact that different $^{12}$CO outflow blobs are coincident with discrete shock-induced H$_2$ knots are all consistent with this being an episodic outflow, similar to the HH 315 molecular outflow \citep{Arc02a,Arc02b}.

We calculate masses of the SE (blue) and NW (red) outflow lobes to be 1 \msun \ and 0.7 \msun , spanning distances of 2.0\arcmin \ (0.14 pc) and 2.7\arcmin \ (0.19 pc) and areas of 2.0 and 2.2 square arcmin ($\sim0.01$ pc$^2$) on the plane of the sky, respectively.  It has previously been noted that additional outflows may exist near this region \citep[e.g.][]{Dav08,Kne00,Cur10b}, but may be confused by the stronger SE-NW outflow.  In \S \ref{sec:cand_svs13} we describe emission which may pertain to an outflow driven approximately south-north by SVS 13C.

\subsubsection{IRAS 2 Outflows} \label{sec:iras2_out}

%%%%%%%%%%%
%fig6
\begin{figure*}[!ht]
\includegraphics[width=\linewidth]{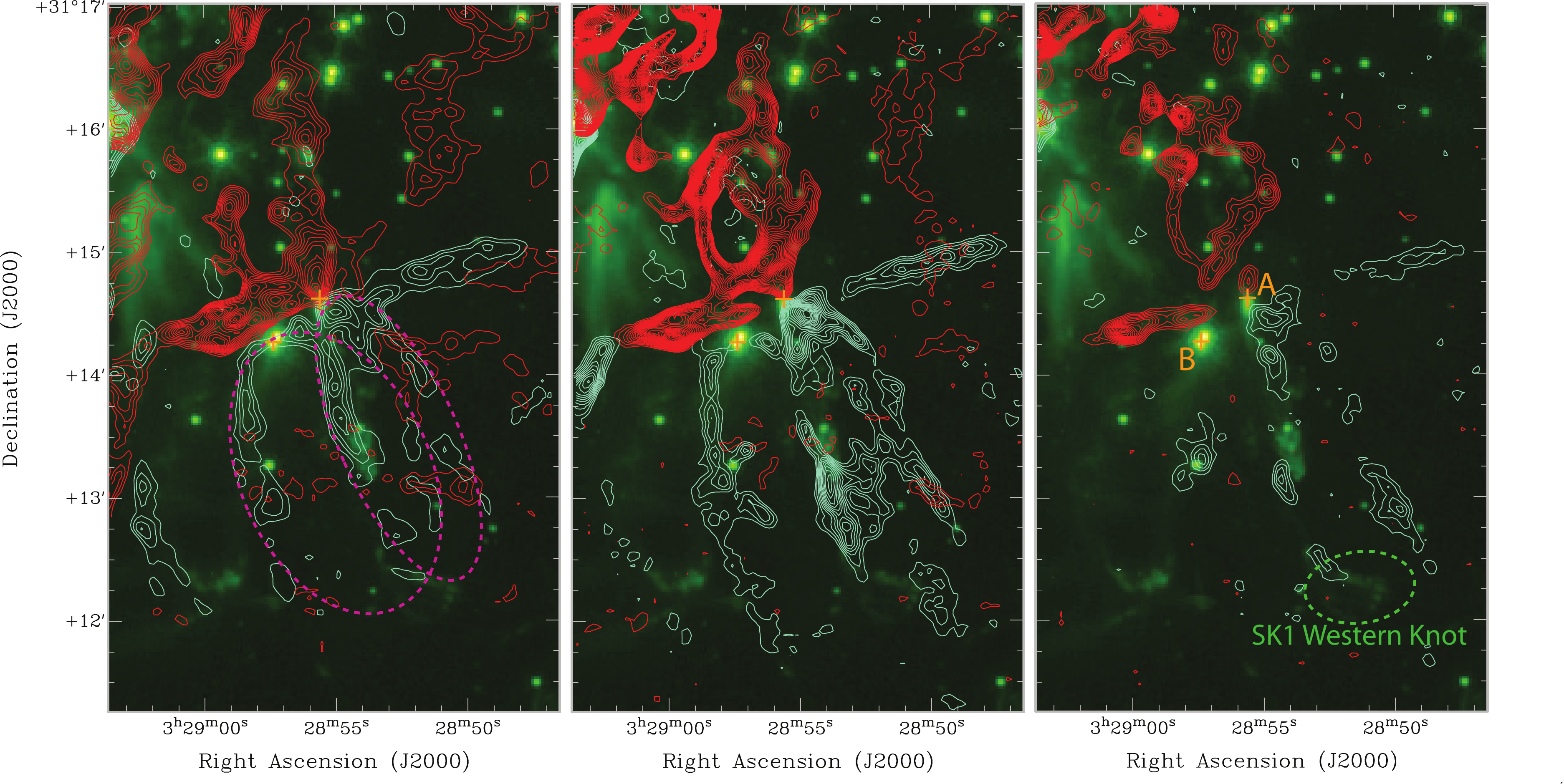}
\caption{Region II low-, mid-, and high-velocity outflows (left, middle, and right panels, respectively).  Blue- and red-shifted CO outflows are shown with blue and red contours, respectively, and background is IRAC 4.5$\mu$m emission \citep{Gut08}.  Continuum sources IRAS 2A and IRAS 2B are marked with plus signs and labeled only in the right panel. Magenta ellipses show the proposed morphology for the southern lobes of the outflows driven by sources IRAS 2B and IRAS 2A (i.e., the lobes IRAS 2B-south and IRAS 2A-south).  Green dashed ellipse marks the region where emission overlaps with the SK1 western knot (see Figure \ref{fig:candsk1}). We group channels as described at the beginning of \S \ref{sec:molecules}.  Contours begin with $3 \sigma$ and increment by $3 \sigma$ for each CO integrated intensity map. }
\label{fig:iras2}
\end{figure*}
%%%%%%%%%%%

We detect three outflows in the IRAS 2 region, where the Class 0 source IRAS 2A and the Class I source IRAS 2B reside (see Figure \ref{fig:iras2}).  Of the outflows associated with IRAS 2A, one outflow is very collimated and oriented approximately west-east (hereafter IRAS 2A west-east), and the other shows a wide-angle morphology oriented approximately south-north (hereafter IRAS 2A south-north).  The shell-like IRAS 2A south-north outflow was previously observed by \citet{Lis88}, and the ``jet-like'' IRAS 2A west-east outflow was observed by \citet{San94} and \citet{Bac98A}.  \citet{Kne00} suggested that IRAS 2A drives both, nearly orthogonal outflows.  \citet{Cur10b} also detect clear orthogonal flows, with IRAS 2A south-north appearing to overlap in the north with the SVS 13A outflow and in the south it may extend beyond the region that we have mapped here.

It has been more recently proposed that IRAS 2A may in fact be a binary \citep{Che13}, and we therefore suggest that the nearly perpendicular outflows that we resolve are being driven by the two components of this binary.  In similar cases where other quadrupolar outflows have been found, subsequent observations with sufficient resolution have revealed that the central protostellar source is a binary, with each component driving a bipolar outflow (e.g., L723, Launhardt 2004; HH111, Reipurth et al. 1999;  CG30, Chen et al. 2008)\nocite{Lau04,Rei99,Che08}. 

Characteristics of the IRAS 2A outflows are given in Tables \ref{tab:areavol} and \ref{tab:sources}.   The three velocity regimes (low-, medium-, and high-velocity for both blue-shifted and red-shifted outflowing gas) are shown in Figure \ref{fig:iras2}, and we find that for IRAS 2A west-east the high-velocity outflowing gas is more collimated than the low-velocity gas.  Specifically, we find that in IRAS 2A west-east, the high-velocity blue- and red-shifted outflow lobes at their widest points have widths of $\sim 16\arcsec$, whereas the low-velocity outflow lobes have widths of  $\sim 22\arcsec$ ($\sim 0.018-0.025$ pc).  \citet{Gue99} see similar morphology in the source HH 211, where high-velocity CO emission shows a collimated jet-like structure and low-velocity emission traces outflow cavities.  The morphology of this outflow is in agreement with the jet-driven outflow paradigm, with low-velocity CO cavities present in the wake of shocks imparted by the high-velocity protostellar jet.  The relatively small overlap of blue- and red-lobes near the driving source IRAS 2A suggests that we are observing IRAS 2A west-east approximately in the plane of the sky with low inclination angle. 

The IRAS 2A south-north outflow, approximately perpendicular to the IRAS 2A west-east outflow, is composed of a blue-shifted (IRAS 2A-south) lobe extending $188\arcsec$ (0.21 pc) south of IRAS 2A and a red-shifted (IRAS 2A-north) lobe extending $133\arcsec$ (0.15 pc) north.  The blue IRAS 2A-south has an opening angle of $\sim 56 \dg$, seen most clearly in low- and mid-velocity outflow emission near IRAS 2A.  At $\sim 50\arcsec$ ($\sim0.06$ pc) south of IRAS 2A, the gas shows nearly parallel cavity walls separated by $\sim 70\arcsec$ ($\sim0.08$ pc). Similar morphology has been seen for the RNO 91 outflow mapped by Lee et al. (2002) and Lee \& Ho (2005)\nocite{Lee02,Lee05}.  In some channels and pixels, emission of IRAS 2A-south may be confused with a southern, comparably widespread outflow lobe driven by IRAS 2B to the east. In particular, in the low- and mid-velocity channels the eastern cavity wall of IRAS 2A-south may be seen projected between the eastern and western cavity walls of IRAS 2B, allowing us to make a tentative distinction of the morphologies (see proposed morphology outlined in Figure \ref{fig:iras2}).  Another confusion comes from SK 1 (see \S \ref{sec:candidates}), which overlaps with IRAS 2A-south in the southern region of our map (see Figures \ref{fig:iras2} and \ref{fig:candsk1}).  Since in this overlap region most emission appears morphologically related to SK 1 (i.e. position angle, elongation approximately west-east rather than south-north) we associate the emission with SK 1 and subtract the mass, momentum, energy and area of SK 1 from that of IRAS 2A-south to determine the values reported in Tables \ref{tab:areavol} and \ref{tab:sources}.

Close to the source, IRAS 2A-north also shows a wide opening angle ($\sim 93 \dg$), particularly at low velocities (see left panel of Figure \ref{fig:iras2}).  Farther from the source, at $\sim 60 \arcsec$ ($\sim 0.07$ pc) north of the driving source, IRAS 2A-north has cavity walls as wide as $90\arcsec$ (0.10 pc).  The extent to which we can identify outflow emission associated with IRAS 2A to the north is limited by ``contamination'' from red-shifted outflow emission associated with SVS 13.  \citet{Kne00} suggested that the IRAS 2A south-north outflow, which they call IRAS 2 NNE-SSW, may have blue-shifted emission extending at least $4\arcmin$ (0.27 pc) to the south, and therefore it is possible that in spite of confusion with surrounding, strong outflows, the red-shifted lobe may extend an equal distance to the north. Further, \citet{Kne00} claim that IRAS 2 NNE extends north beyond the red lobe of SVS 13, and they calculate more mass associated with IRAS 2 NNE to the north of SVS 13 than to the south.  
 
It is apparent in Figure \ref{fig:iras2} that the morphology of IRAS 2A south-north is distinct from that of IRAS 2A west-east.  IRAS 2A south-north spans approximately 5 times more area on the plane of the sky than IRAS 2A west-east, and we measure about 7 times more mass and energy associated with IRAS 2A south-north (see Tables \ref{tab:areavol} and \ref{tab:sources}).  Since IRAS 2A south-north is less collimated than IRAS 2A west-east, and the outflow lobes of IRAS 2A south-north also extend farther so that they are confused with other outflows in our map, several of the characteristics of IRAS 2A south-north are more difficult to determine. 
 
The third outflow in the region, which we tentatively associate with the source IRAS 2B mentioned above, has a clumpy morphology that is further confused by the strong nearby IRAS 2A outflows, and some or all of this clumpy structure may in fact be driven by IRAS 2A, or the candidate outflow SVS 13C-south, described in \S \ref{sec:cand_svs13}.  \citet{Kne00} detect faint, tentative outflow emission extending about 1\arcmin\ south of IRAS 2B, while the map of \citet{Eng99} shows blue-shifted emission $2\farcmin5$ south and red-shifted emission $1\farcmin5$ north that may be associated with IRAS 2B.   

Figure \ref{fig:iras2} shows our suggested outflow morphology, with similar wide angle outflows emanating south from IRAS 2B (hereafter IRAS 2B-south) as in the case of IRAS 2A-south. The opening angle of IRAS 2B-south is $\sim 60 \dg$, seen most clearly in low-velocity channels with $v_{out}<7$ \kms, and beyond $\sim 50\arcsec$ ($\sim0.06$ pc) south of IRAS 2B the cavity walls are nearly parallel, similar to the morphology of IRAS 2A-south.  The proposed northern lobe, IRAS 2B-north, is confused with IRAS 2A-north and the outflow from SVS 13, but particularly in low-velocity channels, emission close to the IRAS 2B continuum source suggests that this source is in fact driving an outflow to the north.  Because we cannot disentangle the outflow from IRAS 2B and the outflows from IRAS 2A and SVS 13C, we do not include a mass estimate for the IRAS 2B outflow.  We acknowledge that the mass estimates of the IRAS 2A and SVS 13C outflows may be over-estimated due to some contribution from IRAS 2B, however our total outflow mass estimate for the region includes the emission pertaining to the IRAS 2B, IRAS 2A and SVS 13C outflows.

%%%%%%%%%%%%%%%%%%%%%%%%%%%%%%%%%%%%%%%%%%%
\subsubsection{IRAS 4 Outflows}\label{sec:iras4_out}

%%%%%%%%%%%%%%%%%%%%
%fig7
\begin{figure*}[!ht]
\includegraphics[width=\linewidth]{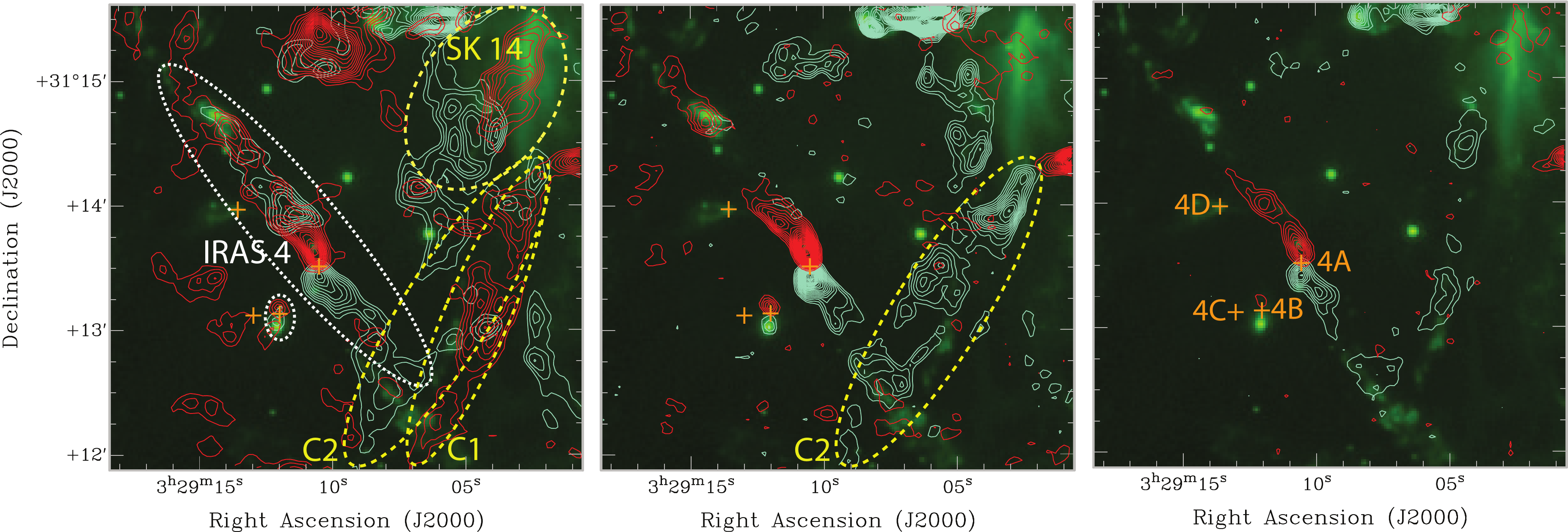}
\caption{Region III low-, mid-, and high-velocity outflows (left, middle, and right panel, respectively).  Blue- and red-shifted CO outflows are shown with blue and red contours, respectively, and background is IRAC 4.5$\mu$m emission \citep{Gut08}.  Continuum sources IRAS 4A, IRAS 4B, IRAS 4C and IRAS 4D are marked with plus signs and labeled in the right panel.  IRAS 4A and IRAS 4B drive outflows outlined approximately with white dashed ellipses.  Southeast of IRAS 4A, IRAS 4B drives an outflow nearly along the line of sight; IRAS 4C and IRAS 4D show no outflow emission.  In the west, we mark candidate outflows with yellow ellipses.  In the southwest, the candidate outflows C1 and C2 are seen most predominantly in low-velocity red-shifted and mid-velocity blue-shifted emission, respectively.  In the northwest, SK14 drives an outflow shown in more detail in Figure \ref{fig:candsvs13c}.  We group channels as described at the beginning of \S \ref{sec:molecules}.  Contours begin with $3 \sigma$ and increment by $3 \sigma$ for each CO integrated intensity map.  }
\label{fig:iras4}
\end{figure*}
%%%%%%%%%%%%%%%%%%%%

Among the Class 0 continuum sources in the IRAS 4 region, described in \S \ref{sec:IRAS4_cont}, the two brightest continuum sources IRAS 4A and IRAS 4B are driving outflows, while IRAS 4C and IRAS 4D appear to have no associated outflow emission. In a detailed molecular line survey, \citet{Bla95} associate outflow emission with IRAS 4A and IRAS 4B, with IRAS 4A driving a highly-collimated jet-like bipolar outflow, and IRAS 4B driving a more compact outflow evidenced by high-velocity wings in the spectrum at the position of the source.  \citet[][see their Figure A3]{Cur10b} detects CO outflow emission associated with both IRAS 4A and IRAS 4B, with the red outflow lobes associated with the two sources resolved at both high- and low-velocities, but the blue lobe of IRAS 4B  does not appear in their low-velocity map and appears unresolved in their high-velocity map.  We detect outflow emission associated with IRAS 4A and IRAS 4B to have distinct morphologies, as suggested by the previously mentioned observations, and in our maps the outflows associated with the two sources are resolved at both low- and high-velocities.

As seen in our map shown in Figure \ref{fig:iras4}, IRAS 4A drives a very collimated outflow, with the blue-shifted lobe towards the south-southwest (hereafter IRAS 4A-south) and the (mostly) redshifted lobe towards the north-northeast (hereafter IRAS 4A-north). The maximum extents of the blue and red lobes that we measure are 104\arcsec (0.12 pc) and  125\arcsec (0.14 pc), respectively. The extent is large enough that both south and north lobes appear to be confused with other outflow emission in the southwest and the north, making it difficult to determine their full extent.  Further, the north lobe extends to the edge of our map in the east.  We note that we detect outflow emission associated with IRAS 4A at all velocity channels with $|v_{out}| > 2$ \kms, even at the outermost channels in our band. In fact, \citet{Cur10b} show that for this source, there is faint outflow emission out to $|v_{out}| = 20$ \kms\ (see the spectra in their Figure B1), and hence our estimates for the mass, momentum and kinetic energy of this outflow should be considered only as lower limits.

Within $\sim 20 \arcsec$ of IRAS 4A, and most notably in the high- and mid-velocity maps, outflow lobes appear to extend nearly north-south, but beyond this distance we find different position angles with decreasing velocities.  Beyond  $20 \arcsec$ we measure position angles of $57 \pm 1 \dg$, $51 \pm 2 \dg$, and $43 \pm 3 \dg$ for the low-, mid-, and high-velocity outflows, respectively.   Fitting an ellipse to the entire outflow and over all velocities, we find a mean position angle of $\sim40$\dg. The blue- and red-shifted lobes intersect but only barely overlap, so the outflows appear to be oriented nearly in the plane of the sky.  We calculate a total outflow mass of 0.3 \msun \ associated with IRAS 4A.  However, we note in \S \ref{sec:candidates} that the candidate outflow lobe C2 likely crosses IRAS 4A-south, and we therefore consider that the mass of IRAS 4A-south may be slightly over-estimated.

\citet{Cho01} observed HCN as a tracer of the outflows in the IRAS 4 region, and they found compact knots associated with IRAS 4A that follow a ``wiggle'' pattern rather than forming a straight trajectory from the driving source.  Later, based on an SiO (1-0) map, \citet{Cho05} measured a change of position angle of the northern collimated lobe, which they called a bend, of about $34\dg$.  The location of this bend (about $25\arcsec$ north of IRAS 4A) is in agreement with the morphology we see in our map and in the CO (2-1) map by \citet{Gir99}.  The CO (2-1) reveals a change in PA of about  $45\dg$, which is closer to the mean PA that we measure.  

Several conclusions can be drawn from the collimated yet ``wiggly/bent'' outflow from IRAS 4A.  \citet{Cho01} suggest several mechanisms for directional and intensity variability, although some of the suggested mechanisms only account for one or the other type of variability, and a combination of mechanisms is likely necessary.  Precession and episodic ejection together adequately account for both the ``wiggles/bends'' (directional variability) and knots (intensity variability) seen in IRAS 4A.  We observe that position angles for both the blue- and red-shifted lobes are consistent, and in particular that gas farther from the source has the largest position angle, so the outflow axis appears to be precessing in the clockwise direction.  The precession may be explained by the fact that IRAS 4A is a binary (see \S \ref{sec:IRAS4_cont}).   \citet{Ter99} suggest that the position axis of an outflow may precess due to the precession of the source, which itself could be caused by the tidal interaction between the circumstellar disk of the outflow source and one (or many) stellar companion(s).  

It can also be seen in Figure \ref{fig:iras4} that the northern lobe of the IRAS 4A outflow consists of blue- and red-shifted outflow emission (specifically at low-velocities), whereas the southern lobe is almost entirely blue-shifted.  In particular, in IRAS 4A-north we detect high-velocity red-shifted emission spatially coincident with lower-velocity blue-shifted emission (in channels with $v_{LSR}=3-17$ \kms). The blue-shifted emission has a bow-shaped morphology, and it is coincident with the position of the bend in the northern lobe.  Since the blue- and red-shifted emission follow a continuous outflow trajectory to the north of IRAS 4A, we suggest that both components comprise the same northern outflow lobe, and the spatial coincidence of the red- and blue-shifted outflow emission is consistent with the scenario of precession about an axis approximately parallel to the plane of the sky.

The observed IRAS 4B outflow kinematics and morphology differ drastically from IRAS 4A.  Although it is also driven by a Class 0 source, IRAS 4B shows weaker outflow emission than IRAS 4A, and a more substantial fraction of the emission is low- and mid-velocity.  In contrast to IRAS 4A, the IRAS 4B outflow appears oriented along the line of sight and it is very compact as seen in Figure \ref{fig:iras4}.  Possible projection scenarios for the IRAS 4B outflow relative to that from IRAS 4A are presented by \citet{Yil12} based on multi-transition observations of CO and several of its isotopologues, with models assuming that the intrinsic lengths of the outflows are similar.  We calculate a mass of 0.02 \msun \ for IRAS 4B, with each lobe only extending about 10\arcsec \ on the plane of the sky, but spanning about $v_{cloud}\pm 5$ \kms \ in radial velocity. This extent is in agreement with those reported by \citet{Yil12} based on CO (3-2) and CO (6-5) observations made with JCMT and APEX, although in those observations the outflow lobes were reported as unresolved.

As stated previously, we detect continuum emission attributed to both binary components IRAS 4B and IRAS 4C, but no outflow emission associated with IRAS 4C was detected.  Similarly we do not detect outflow emission associated with the weaker continuum source IRAS 4D to the northeast.

%%%%%%%%%%%%%%%%%%%%%%%%%%%%%%%%%%%%%%%%%%%
\subsection{Outflow Candidates} \label{sec:cand_out}

Here we present several outflow candidates based on $^{12}$CO emission morphology that were previously unidentified or confused with stronger surrounding outflows.  We associate candidate CO outflow emission with the protostellar sources SVS 13C, SK 14, and SK 1. The candidate outflows that we cannot associate with a driving source are labeled C1, C2, C3 and C4.

\subsubsection{SVS 13C}\label{sec:cand_svs13}

%%%%%%%%%%%%%%
 %fig8
\begin{figure}[!ht]
\includegraphics[angle=0,width=\linewidth]{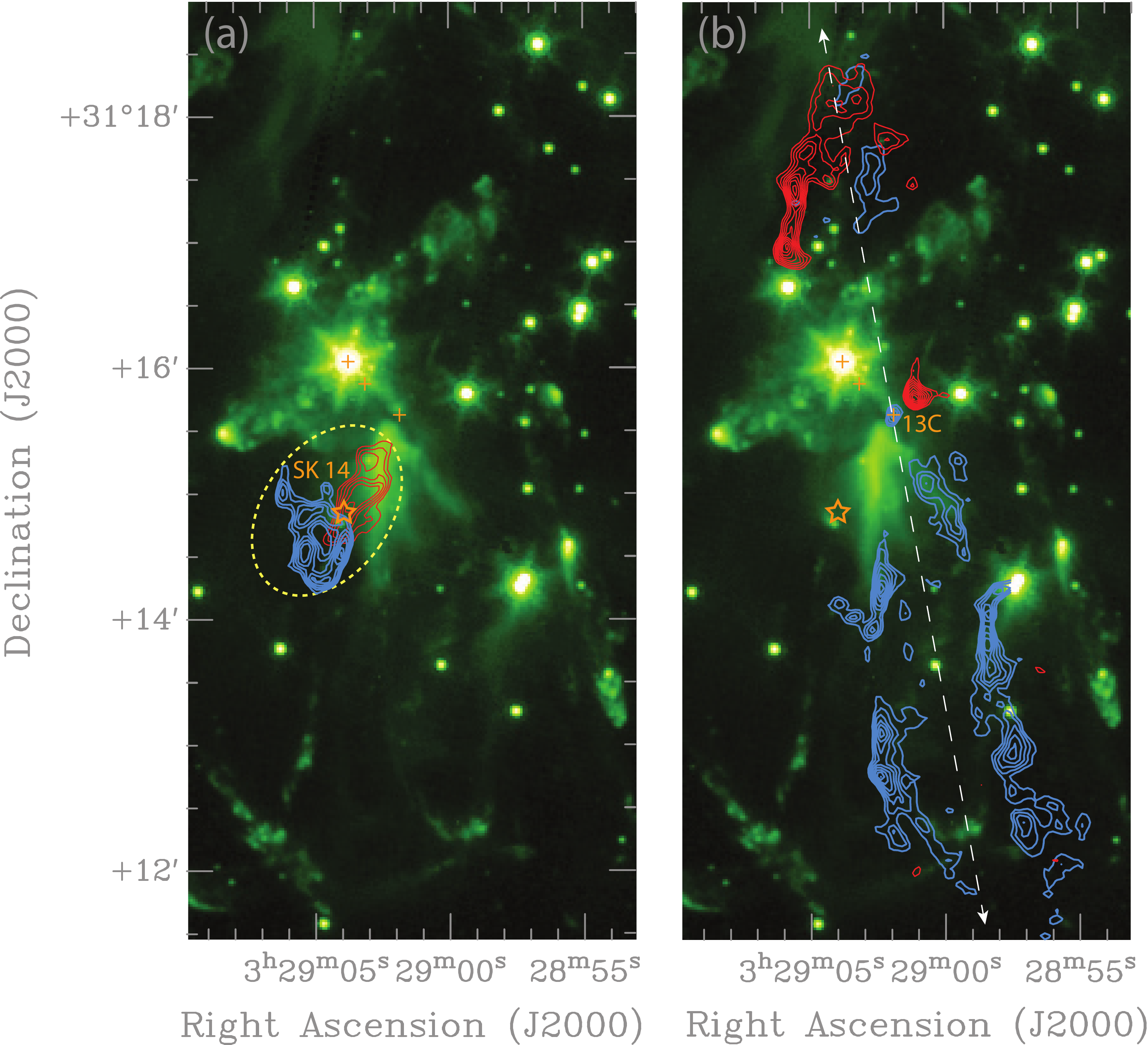}
\caption{Candidate outflows associated with SK 14 (panel a) and SVS 13C (panel b), corresponding to Region IV in Figure \ref{fig:wholemap}.  (a) Contours are integrated intensity for channels with $v_{LSR}=-1$ to $6$ \kms\ (blue) and  $v_{LSR}=9$ to $11$ \kms\ (red), beginning with $3\sigma$ and incrementing by $\sigma$, where $\sigma$ is the rms of the respective integrated intensity maps.  (b) Contours are integrated intensity for channels with $v_{LSR}=4$ to $6$ \kms\ (blue) and  $v_{LSR}=12$ to $14$ \kms\ (red), beginning with $3\sigma$ and incrementing by $3\sigma$.  For clarity in each panel of this figure, we include only contours pertaining to the outflows SK 14 and SVS 13C, and we erased the contours of outflow emission from other nearby sources (e.g., SVS 13A).  Crosses mark the continuum sources in the SVS 13 region, and the star symbol marks SK 14, for which we did not detect continuum emission.  The dashed white line in (b) marks an axis for the SVS 13C north-south outflow with position angle 8\dg, fit by eye.  Arrows suggest that the outflow lobes may extend beyond the edge of our map.  Background image is IRAC 4.5 $\mu$m emission.  }
\label{fig:candsvs13c}
\end{figure}
%%%%%%%%%%%%%%
%%%%%%%%%%%%%%
%fig9
\begin{figure*}[!ht]
\includegraphics[angle=0,width=0.8\linewidth]{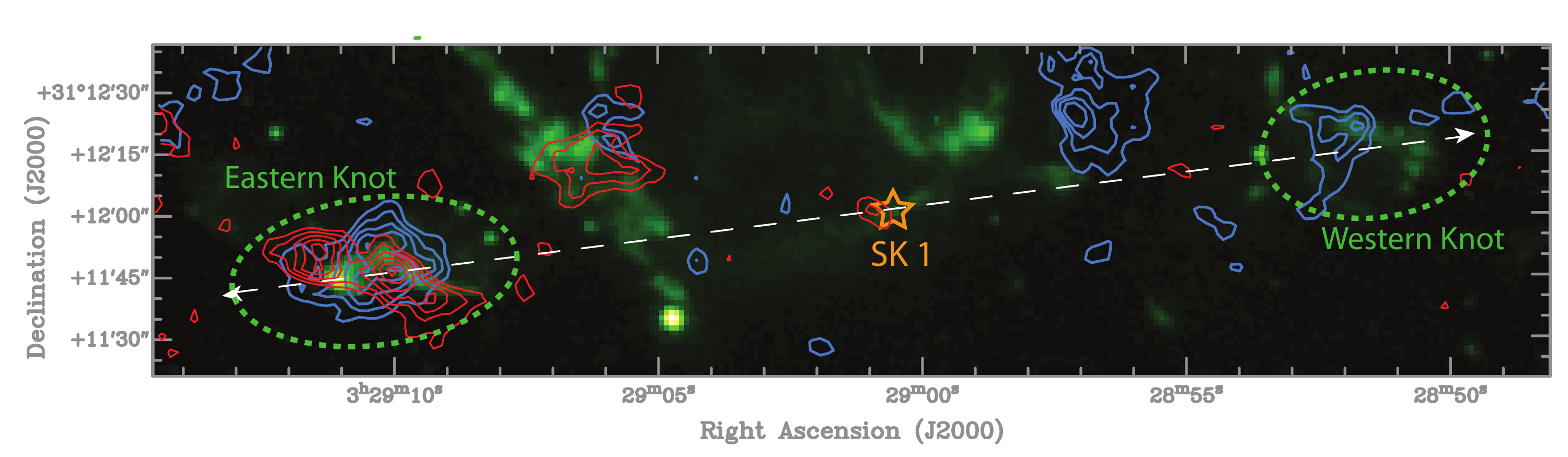}
\caption{Outflow associated with SK1 in the southern region of our map, corresponding to Region V in Figure \ref{fig:wholemap}.  Contours show integrated intensity for channels with velocities $v_{LSR}=5.0$ to $6.5$ \kms (blue) and  $v_{LSR}=10.02$ to $12.2$ \kms (red), beginning with $3\sigma$ and incrementing by $2\sigma$, where $\sigma$ is the rms of the respective integrated intensity maps.  The outflow emission detected in our map is comprised of knots in the east and west, marked with green dashed ellipses. The dashed white line marks an axis for the outflow with position angle 97\dg.  Background is IRAC 4.5 $\mu$m emission, and it is evident that in the west and the east CO outflow emission is coincident with bow-shocks.  }
\label{fig:candsk1}
\end{figure*}
%%%%%%%%%%%%%%

In \S \ref{sec:svs13_cont}, we described three continuum sources detected in the SVS 13 region.  Of these, SVS 13A is the continuum source associated with the IR source SVS 13 driving the bipolar HH 7-11 outflow, which we described in \S \ref{sec:svs13_out}.  In addition to this outflow, we also detect molecular outflow emission extending south-north on the plane of the sky that appears to be associated with the southern-most source in the group, SVS 13C (Region IV in Figure \ref{fig:wholemap}, see also Figure \ref{fig:candsvs13c}).  Outflow emission south of SVS 13 has previously been observed \citep[e.g.][]{Kne00,Dav08,Cur10b}, but the driving source has not been conclusively or consistently identified.  \citet{Kne00} associate this emission with an outflow driven by SVS 13B, but they also acknowledge that a dust shell found south of SVS 13 \citep{Lef98,San01} may have been created by SVS 13B or SVS 13C (they use the name H$_2$O(B) for this source).

Here we provide more evidence that this outflow is driven by SVS 13C, and we propose a northern outflow lobe counterpart.  South and north of SVS 13C we detect CO emission up to $|v_{out}|=9$ \kms  \ along an axis with position angle 8\dg, pertaining to blue and red lobes extending $\sim 212$\arcsec \ (0.24 pc) south and $\sim 180$\arcsec \  (0.21 pc) north, respectively (hereafter SVS 13C-south and SVS 13C-north).  

Particularly in SVS 13C-south, CO emission approximately follows the nebula seen in 4.5 $\mu$m emission, shown in Figure \ref{fig:candsvs13c}.  In the far south CO outflow emission is coincident with 4.5 $\mu$m emission that has a bow-shaped morphology and which may be a shock from the SVS 13C-south flow.  Proper motions for this structure are up to 100 \kms \ in the southern direction \citep{Rag13}, and such high velocity and large spatial extent on the plane of the sky suggest a low inclination of this outflow lobe with respect to the plane of the sky.  Some of the emission detected in the southwestern component of SVS 13C-south may also be contributed by the IRAS 2B outflow (see Fig. \ref{fig:iras2}), but as can be seen in Figure \ref{fig:candsvs13c} the emission is consistent spatially with an outflow lobe driven by SVS 13C.

To the north of SVS 13C, we see strong emission comprising the red outflow lobe SVS 13C-north which counters the blue outflow lobe SVS 13C-south, but which is also confused with the northwest red SVS 13A outflow lobe described in \S \ref{sec:svs13_out}, or possibly the northern outflow lobe driven by IRAS 2 \citep{Kne00}.  Another possibility is that this emission may be associated with a source at or beyond the northern edge of our map, further reason to extend the map to the north in order to more conclusively identify the counterpart lobes.  However, based on our map, we suggest that SVS 13C-south and SVS 13C-north are the blue and red lobes of the molecular outflow associated with SVS 13C.  For clarity, in Figure \ref{fig:candsvs13c} we show the SVS 13 region with contours only pertaining to emission which we associate with SVS 13C but not the SVS 13A SE-NW outflow.  Since both lobes of the SVS 13C outflow shown in Figure \ref{fig:candsvs13c} extend out to the edge of our map, it is very likely that these lobes extend even farther than what we map here. 

Near the position of the SVS 13C continuum source are two concentrated emission blobs in  blue- and red-shifted channels up to $|v_{out}|=5$ \kms.  The blue-shifted blob is coincident with the position of SVS 13C, and the red-shifted blob extends northwest only $\sim10-20 \arcsec$ (projected on the plane of the sky) from SVS 13C, reminiscent of the IRAS 4B outflow.  One possible explanation is that these two blobs represent the lobes of an outflow that is oriented nearly along the line of sight (similar to the IRAS 4B outflow) and perpendicular to the main north-south outflow driven by SVS 13C.  If this is the case, then SVS 13C is likely a binary source driving two bipolar outflows. Alternatively, the blobs may be part of the wall of the SVS 13C-north outflow lobe.  Further observations are needed to explain this emission structure.

\subsubsection{SK 14 Candidate Outflow}

In the region southeast of SVS 13 and northwest of IRAS 4, we detect a bipolar outflow oriented southeast-northwest seen particularly in low-velocity channels with $|v_{out}|=2$ to $4$ \kms, but also with blue-shifted emission at higher velocities (see Figure \ref{fig:candsvs13c}, also Figure \ref{fig:iras4}).  The outflow lobes intersect at the location of the Class 0 source SK 14 detected by \citet{San01}, which in addition to sub-mm detections \citep{San01,Hat05} also has counterparts at radio \citep{Rod99}, millimeter \citep{Eno06} and IR \citep{Gut08,Eva09} wavelengths.  As indicated in Table \ref{tab:ysos}, this is a Class 0 source from SED-based indicators \citep[e.g.][]{Hat07a}.  In previous outflow studies, such as that of \citet[][see their Table 1]{Hat07b} who observed CO (3-2) with JCMT, the outflow associated with this source is suggested, but has been confused by nearby stronger outflow emission, particularly emission related to SVS 13.  The spectral and spatial resolution of our observations allow us for the first time to distinguish the velocity structure and morphology of the outflow lobes.  This also has implications for distinguishing the candidate outflow lobes C1 and C2, described in \S \ref{sec:candidates}.

We detect emission associated with SK 14 in blue channels with velocities up to $|v_{out}|=9$ \kms, whereas emission in the red channels is only seen up to $|v_{out}|=3$ \kms.  At low velocities, the blue outflow has a donut-shaped morphology and ``clumpy'' emission similar to that of SVS 13, although the structure is smaller and the integrated emission is not as strong.  The red-shifted outflow emission remains narrower and weaker than the blue emission, and the northern extent of the outflow lobe may be confused with SVS 13 and other strong outflows to the north and east.  We measure a position angle of 145\dg, and blue- and red-shifted outflow lobe extents of 71\arcsec (0.08 pc) and 58\arcsec (0.07 pc), respectively.

\subsubsection{SK 1 Candidate Outflow}

The source SK 1 \citep{San01}, also identified as Bolo 41 \citep{Eno06} and HRF 65 \citep{Hat07a,Hat09,Cur10b}, lies at the southern edge of our map (Region V of Figure \ref{fig:wholemap}, see also Figure \ref{fig:candsk1}), and has been previously classified as a Class 0 object (see Table \ref{tab:ysos}).  \citet{Cur10b} show that the source drives a bipolar outflow extending east-west nearly $10\arcmin$, with the red lobe to the east and the blue lobe to the west, both extending beyond the edges of our map.  The knots that we detect associated with SK 1 have a bow-shape and are coincident with IRAC 4.5 $\mu$m emission at $\sim2.5\arcmin$ (0.2 pc) from the driving source, shown in Figure \ref{fig:candsk1}.  These features are presumably caused by outflow shocks, characteristic of a young, jet-like outflow.  It is very likely that we only detect these distinct blobs of emission, rather than a contiguous outflow morphology, because this source is in the region of our map with the poorest sensitivity and we are only able to detect the brightest blobs of this outflow. We only include these features (see Figure \ref{fig:candsk1}) in our calculations pertaining to SK 1 in Tables \ref{tab:areavol} and \ref{tab:sources}. 

Based on the fact that the eastern-most knot is seen in both blue- and red-shifted emission, as well as proper motions showing high velocity ($\sim80$ \kms) along the plane of the sky \citep{Rag13}, it is very likely that this outflow has a low inclination angle with respect to the plane of the sky.  The emission that we detect is coincident with previous observations, particularly the $^{12}$CO (J=3-2) emission observed by \citet{Cur10b}, and with our higher-resolution maps (see Figure \ref{fig:candsk1}) we see that the east-west outflow lobes are about $30\arcsec$ wide, narrower than previously determined.  Some of the emission previously attributed to this outflow may in fact be contributed by several distinct outflows in the region.  For example, the blue (west) lobe of SK 1 may be ``contaminated'' by the blue-shifted emission of the southern lobe from SVS 13C or IRAS 2 (discussed above), while in the east we see both red and blue emission that may be associated with other candidate outflows (discussed below).

\subsubsection{Candidate Outflow Lobes Without Identified Sources} \label{sec:candidates}

We identify four candidate outflow lobes within the map based on morphology, but for which we cannot identify conclusively their driving sources.  In the southeast region of our map (within Region III of Figure \ref{fig:wholemap}, shown also in Figure \ref{fig:iras4}) we see a narrow red outflow feature which we name C1 that extends at least 180\arcsec (0.2 pc).  This structure has a position angle of 157\dg, and is coincident with filamentary emission seen in IRAC 4.5 $\mu$m emission \citep{Gut08} along its northeastern edge.  We see emission associated with this feature in red channels with $|v_{out}|\lesssim3$ \kms.  With a relatively small extent in radial velocity, C1 is most apparent when inspecting a moment map of only the low-velocity channels (i.e. left panel of Figure \ref{fig:iras4}, rather than Figure \ref{fig:wholemap}).  Within the extent of our map we do not see any blue-shifted emission that could be conclusively identified as the counter-lobe of C1.  

Also in Region III and marked in Figure \ref{fig:iras4}, slightly northeast of C1 we see a candidate outflow lobe C2 with position angle 148\dg\ and with similar morphology, but with blue-shifted velocities $|v_{out}|\lesssim8$ \kms. The southwestern edge of C2 is coincident in projection with the northeastern edge of C1 and the shocked emission seen in IRAC 4.5 $\mu$m emission.  It is difficult to conclusively associate the shocked emission with either C1 or C2, but the position angle of the IRAC filament appears to be more consistent with C1.  H$_2$ emission knots in this region have been observed in the near- and mid-IR by \citet{Dav08} and \citet{Mar09}, respectively,  with position angle of about 160-170\dg, although they were previously associated with SVS 13B, whose tentative outflow is discussed more below.

The blue-shifted C2 outflow emission extends at least 122\arcsec (0.14 pc), where the southern tip appears to coincide with the blue southwestern edge of IRAS 4A-south.  To the southeast of IRAS 4A-south, there also appears to be emission associated with C2 and with the same position angle, but since we cannot disentangle C2 from IRAS 4A-south, and in the region where the two overlap the majority of the emission appears to be associated with IRAS 4A-south, we only include the emission west of IRAS 4A-south in our calculations for C2, presented in Tables \ref{tab:areavol} and \ref{tab:sources} .  

In previous molecular outflow and H$_2$ observations, an outflow lobe has been tentatively suggested to extend south from SVS 13B, comprising SK 14 and C2 and reaching IRAS 4A-south.  For example, the CO observations of \citet[][see their Figures 2 and 3]{Kne00} and \citet[][see their Figure A3]{Cur10b} reveal blue-shifted emission in the region between SVS 13 and IRAS 4, but upon inspection of the maps in these two studies it is apparent that the emission does not follow a consistent position angle nor does it have the morphology characteristic of other better-defined outflow lobes in the region.  Further, there appears to be a dip in CO intensity along the previously proposed feature, midway between SVS 13 and IRAS 4.  With our higher-resolution maps, having identified the SK 14 outflow (blue- and red-shifted lobes) and revealing the blue-shifted outflow lobe candidate C2 (in addition to the red-shifted lobe C1) we therefore suggest that the observed morphology of the CO emission towards the southeast of SVS 13 and northwest of IRAS 4A-south may instead be explained by several distinct outflow features, for which we cannot conclusively identify the driving source(s).

In Region I in the northeastern corner of our map we detect clumps of red-shifted emission, which we name C3 and C4, shown in Figure \ref{fig:svs13}.  These structures show elongated outflow-like morphologies, with a position angle of $\sim160$\dg, and are coincident with extended 4.5 $\mu$m emission.  C3 extends $\sim200$\arcsec (0.23 pc), with a $\sim50$\arcsec \ blob of emission to the northwest, and a more narrow filamentary structure with nearly parallel walls extending towards the southeast.  C4 is narrower, with length of 144\arcsec (0.16 pc) and width of only  18\arcsec (0.02 pc).  We expect the blue-shifted outflow lobes associated with C3 and C4 to lie beyond the north or east edges of our map.  At least one of these outflow features might be associated with the protostar IRAS 7 (see Table \ref{tab:ysos}) which lies at the northern edge of our map and also drives the HH6 jet and associated outflow.  

%%%%%%%%%%%%%%%%%%%%%%%%%%%%%%%%%%%%%%%%%%%

\subsection{Mass, Momentum and Energy of Outflows}\label{sec:mass}

Here we describe the method for determining mass, momentum and energy of the outflows based on CO emission.  The outflows described in \S \ref{sec:molecules}-\ref{sec:cand_out} were each fit with an ellipse in order to approximate the outflow morphology and delineate emission associated with an individual outflow.  Each ellipse was fit to include the outflow driving source (if identified, with the exception of SK1 explained in footnote of Table \ref{tab:areavol}), as well as the farthest and widest pixels of CO emission greater than 3$\sigma$, and then each velocity channel was inspected by eye to ensure that all emission greater than 3$\sigma$ that was consistent spatially and spectrally with the individual outflow morphology was included.  Parameters of these ellipses, which allowed us to calculate outflow morphology characteristics, are given in Table \ref{tab:areavol}.  

Inspecting the channel maps of the combined CARMA+FCRAO observations, we detect intervening large-scale, smooth CO emission in channels with velocity between 4 and 10 \kms \ likely due to another cloud along the line of sight \citep[see appendix of][]{Arc10}. This extended emission is not seen in CARMA-only maps because it is filtered out by the interferometer.  In the combined interferometer and single dish map, we measured the emission within regions with no outflows, and we found that the intervening emission is approximately uniform in each velocity channel and peaks at $5.9$ \kms.  For our outflow mass calculations, we subtract this excess emission for each channel where significant extended emission is detected. 

We follow the method of \citet{Arc01} -- who followed a modified method of \citet{Bal99} and \citet{Yu99} -- to estimate the outflow mass, using $^{12}$CO and $^{13}$CO data to correct for the velocity-dependent opacity of the $^{12}$CO line. For each outflow lobe we calculate the average $^{13}$CO emission and average $^{12}$CO emission for each 0.3 \kms \ wide channel, and using the average spectra we determine the ratio of the intensity, $R_{12/13}$, within the ellipse used to define the outflow lobe emission.  We fit a second-order polynomial to the ratio, constrained to have a minimum value at the velocity of the ambient cloud ($v_{LSR}=8$ \kms), and excluding velocity channels closest to the ambient velocity with $v_{LSR}=6-10$ \kms .  We truncate the function $R_{12/13}$ at 62, the assumed isotopic ratio \citep{Lan93}.  This parabolic fit is used to extrapolate to the high-velocity wings of the outflow where the $^{13}$CO is too weak to be reliably detected (explained below).

We assume that $^{13}$CO is optically thin at outflow velocities, and if $^{13}$CO emission from a position $(x_i,y_i)$ at a velocity $v_i$ is greater than or equal to three times the rms noise of the spectrum at that position, then we use the intensity for that velocity, $I_{13}(x_i,y_i,v_i)$, to estimate the $^{13}$CO optical depth and column density (from which we obtain the gas mass).    If $^{13}$CO emission is less than three times the rms noise, then we use the $^{12}$CO emission at that position $(x_i,y_i)$ and velocity $v_i$ to estimate $I_{13}(x_i,y_i,v_i)=I_{12}(x_i,y_i,v_i)[R_{12/13}]^{-1}$.

We determine optical depth of the $^{13}$CO line according to the following equation \citep{Wil09}:

\begin{equation}
\tau_{13}(x,y,v)=-\ln\left[1-\frac{I_{13}(x,y,v)}{T_0([\exp(T_0/T_{ex})-1]^{-1}-0.16)} \right]
\end{equation}
where $T_0=h\nu/k=5.29$ for $^{13}$CO, and the excitation temperature is found based on the peak intensity, $I_{peak}$, of $^{12}$CO for each pixel, using the relation
\begin{equation}
T_{ex}=\frac{5.53}{\ln\left[1+5.53/(I_{peak}+0.82) \right]}.
\end{equation}
We found that the excitation temperature is $T_{ex}=20\pm4$ K in our map. At each pixel position and velocity, the column density can be determined according to the following equation \citep{Wil09}:
 
\begin{equation}
N_{13}(x,y,v)=(2.5\times 10^{14}) T_{ex}\frac{\tau_{13}(x,y,v)dv}{1-\exp(-T_0/T_{ex})}.
\end{equation}

The column density summed over all velocity channels is $N_{13}(x,y)=\Sigma_{vel} N_{13}(x,y,v)$.  The outflow (molecular hydrogen) mass at each position pixel is given by $M(x,y)=m_{H_2}N_{H_2}(x,y)A$, where $m_{H_2}$ is the mean molecular weight taking into account the abundance of helium and other trace constituents, in this case 2.72 times the mass of a hydrogen atom, and $N_{H_2}=7\times10^5 N_{13}$ \citep{Fre82} is the molecular hydrogen column density.  The ratio of [H$_2$/$^{13}$CO] is an important source of uncertainty in the calculations.  Some studies found this ratio to be about 0.5 times the value we use here \citep[e.g.][]{Pin08}, so the outflow masses may be less according to this factor, but here we use the value of $7\times10^5$ from \citet{Fre82} to allow for more straight-forward comparisons with other similar studies of the same outflows that we present here.  $A$ is the physical area of the pixel at the distance of the source; in our case each pixel is $2\arcsec\times2$\arcsec, which corresponds to $5\times10^{31}$ cm$^2$.  Finally, the mass summed over the outflow area is $M=\sum_{area}M(x,y)$.

Outflow momentum ($P_{out}$) and energy ($E_{out}$) are found according to the following relations, respectively:

\begin{eqnarray}
P_{out}&=&\Sigma_{vel} M(v)|v-v_{cloud}|\\
E_{out}&=&\frac{1}{2} \Sigma_{vel}M(v)|v-v_{cloud}|^{2},
\end{eqnarray}
where $M(v)$ is the outflow lobe mass for a given velocity channel, $v$ is the velocity corresponding to that channel, and $v_{cloud}=8$ \kms \ is the cloud velocity.  We sum over all velocity channels with $|v-v_{cloud}|> 2$ \kms.  Characteristics of outflows associated with known protostellar sources and candidate outflows are itemized in Table \ref{tab:sources}, where we have made no correction for inclination of the outflow with respect to the plane of the sky.  

In total, we measure 6 \msun \ of gas that we associate with outflow lobes or outflow candidates.  One source of uncertainty that propagates in our mass estimate, and consequently momentum and energy, is the excitation temperature.  The approach described here allows for non-constant excitation temperature throughout the region mapped, calculating excitation temperature pixel-by-pixel.  As mentioned previously, we find a mean excitation temperature of 20 K.  If we assume a constant $T_{ex}=20$ K,  we calculate masses for each outflow that are $95 \pm 6$ \% of those calculated following the pixel-by-pixel method described previously.  Outflows typically have $T_{ex}\sim10-50$ K, and we follow several studies which suggest that the outflows in Perseus, and NGC 1333 in particular, likely have temperatures at the lower limit of this range \citep[e.g.][]{Kne00}.   Other studies of the NGC 1333 region \citep[e.g.][and references therein]{Hat07b,Cur10b} used a higher excitation temperature $T_{ex}=50$ K, acknowledging $T_{ex}$ up to 100 K.  Using a constant $T_{ex}=50$ K, we calculate outflow masses approximately twice those reported here.  

We acknowledge that within the $49$ square arcmin (0.23 pc$^2$) area of the map, there are a few regions where we detect CO emission with a signal-to-noise greater than 3, which we cannot associate with any  of the identified or candidate outflows presented here.  It is possible that some outflows are un-identifiable due to resolution or in the case that some YSOs beyond the edges of our map are driving outflows which extend into the region which we mapped.  We suspect that with improved resolution, and by mapping a larger surrounding area, we will be able to better associate and quantify the outflow emission in this region.

We calculate the total momentum and energy for identified outflows and outflow candidates to be $19$ \msun \kms \ and $7.2\times10^{44}$ erg, respectively.  Assuming an average inclination angle with respect to the line of sight $\xi=57.3\dg$ \citep[following][]{Bon96,Nak11} we estimate the total outflow momentum and energy (corrected by outflow inclination angle) to be $35$ \msun \kms \ and $2.5\times10^{45}$ erg, respectively.

%%%%%%%%%%%%%%%%%%%%%%%%%%%%%%%%%%%%%%%%%%%%%%%
\section{Discussion} \label{sec:discussion}

\subsection{Area and volume occupied by outflow emission}

From Figure \ref{fig:wholemap} we can see that CO outflow emission covers much of the area mapped with highest sensitivity (see also Figure \ref{fig:gut08}). At first glance this seems to indicate that outflows in this region could have a major impact on the environment.  In order to quantify this, we estimated the percentage of the area of the cloud in the mapped region that corresponds to outflows.  We detect CO emission with $|v_{out}|>2$ \kms \ and with a signal to noise greater than three in 70\% of the area we mapped projected on the plane of the sky.  This two-dimensional diagnostic gives a sense that outflows occupy a major fraction of the cluster, and as a further diagnostic of outflows within the three-dimensional cluster, we also compare outflow volume with cloud volume according to the following method.  

We estimate the volume of each outflow based on the ellipses we fit to the emission associated with identified outflows and outflow candidates described in \S  \ref{sec:molecules} (see Table \ref{tab:areavol}).  We assumed that the outflow lobes in three dimensions are approximately ellipsoidal such that the depth is equal to the minor axis of the two-dimensional ellipse on the plane of the sky.  Without correcting for inclination in the plane of the sky, our estimate is a lower limit to the outflow volume, the most drastic case being IRAS 4B (see \S \ref{sec:iras4_out}) which is likely oriented along the line of sight, and therefore whose depth is likely substantially larger than its width on the plane of the sky.  Further, we re-iterate that unlike the calculation of area occupied by outflows for which we include all CO emission with $|v_{out}|>2$ \kms \ that we detect in the region we mapped, in the calculation of outflow volume we measure only outflows (and outflow candidates) that we explicitly identify in the region and fit with ellipsoids. 

We estimate the volume of the cloud within the region mapped by our CARMA observations, which we assume to be the approximate center and most active region of the cluster, to be $0.1-0.3$ pc$^3$ based on the following assumptions.  We assume that the depth of the cloud in this region is equal to $D=0.8-2$ pc ($12^\prime-30^\prime$ at a distance of 235 pc), corresponding to the minor axis of the cloud determined from C$^{18}$O observations \citep{Rid03} and $^{13}$CO observations \citep{Arc10}, respectively.  We approximate the volume of the cloud within the mapped region to be an ellipsoid such that $V=\frac{4}{3}\left(A\right) \left(\frac{D}{2}\right)$, where $A=0.23$ pc$^2$ is the area of our map.  Including ellipsoids fit to all the identified outflows and outflow candidates (see Table \ref{tab:areavol}), we calculate that in the region we mapped the volume of the cloud occupied by outflows is $\sim2-6$\%.  

This volume filling factor is much less than the area filling factor, leading us to conclude that even in the very active central region of the cluster, the majority of the volume is not occupied by outflows during the early stages of star formation that we observe here.  However, this does not necessarily negate the possibility that outflows impact the surrounding large-scale cluster gas, perhaps with the aid of magnetic fields acting as an important agent that helps sustain outflow-driven turbulence \citep[e.g.][]{Wan10}.  Since most of the sources driving outflows in the region are Class 0, with time the outflows will occupy more of the cloud volume as they become less collimated and more wide-spread. If we assume that all outflows in our map will eventually have the volume of the Class I source SVS 13, then the 22 outflow lobes would potentially occupy up to 30\% of the region's volume in the next $\sim0.5$ Myr, the lifetime for the Class I stage reported by \citet{Eva09}.  

\subsection{Energy imparted by outflows}

IRAC images and our millimeter observations show the prevalence of outflows within this region, implying that indeed these outflows have an impact on the surrounding cloud.  From the discussion in \S \ref{sec:molecules}-\ref{sec:cand_out} and the values listed in Table \ref{tab:sources}, it is apparent that the identified outflows have a range of morphologies and energetics which contribute to the total outflow activity in the region.  In order to quantify the impact of this outflow activity, we need to compare the total outflow energetics with gravitational energy and turbulence within the cloud according to the method described below.  

We calculate the cloud mass within the region mapped by our CARMA observations (see Figure \ref{fig:wholemap}) to be $M_{cl}=140$ \msun, using the FCRAO $^{13}$CO data for the same region and $T_{ex}=13$ K \citep{Arc10}.  We estimate the gravitational binding energy ($W=-GM_{cl}^2/R_{cl}$) to be $7.2\times10^{45}$ ergs in this region.  We assume a cloud radius ($R_{cl}$) of 0.24 pc, which is half the extent of the $7^\prime$ mapped region.  The energy associated with identified outflows (with no inclination angle accounted for) equals about 10\% of the gravitational energy in the region, and about one-third when using an average outflow inclination angle of $\xi=57.3\dg$.  Although outflow energy does not exceed the region's gravitational energy, the outflows that we detect will likely cause mass to escape from the cloud given that the outflows have velocities greater than the escape velocity of $v_{esc}=2.2$ \kms \ for the region \citep{Arc10}.  Given that up to 30\% of the cloud volume will eventually be occupied by outflowing gas (see previous section), and assuming a homogeneous density for the cloud, about 30\% of the mass will be in outflows with the potential to escape.  Since gravitational binding energy is proportional to $M^2$, the gravitational energy will decrease by a factor of $(0.7)^2\approx50$\% during the next $\sim0.5$\ Myr.

We also investigate the impact outflows have on the turbulence of the cloud.  Based on cloud mass $M_{cl}=140$ \msun \ and average $^{13}$CO velocity width (FWHM) $\Delta V=2.2$ \kms \ \citep{Arc10}, we find that $E_{turb}=\frac{3}{16}\ln(2)M_{cloud}\Delta V_{turb}^2=1.8 \times10^{45}$ erg.  This equals about one quarter of the gravitational potential energy for the same region, and is approximately 2.5 times greater than the total energy of identified outflows in the region (without correction for inclination angle). Depending on outflow inclination angle, the outflow energy may be approximately equal to turbulent energy.  Indeed outflows have the potential to contribute energy comparable to the amount of turbulence we measure in the region, and therefore outflows may be an important turbulent agent.

A further diagnostic of outflows and turbulence compares outflow luminosity and turbulent dissipation rate (i.e., the power needed to maintain the turbulence in the cloud), taking into account the outflow and turbulent energies and timescales according to the following method.  We begin by assuming an outflow timescale equal to the mean dynamical time of outflows in the region. Our estimate of mean dynamical time is $\sim5\times10^4$ years, calculated as $t_{dyn}=<l_{out}>/v_{char}$, where $<l_{out}>=0.15$ pc is the mean outflow length from Table \ref{tab:areavol} ($<l_{out}>=0.27$ pc, with $\xi=57.3\dg$), and $v_{char}=P_{out}/M_{out}=3.1$ \kms \ ($v_{char}=5.8$ \kms, with $\xi=57.3\dg$).

In estimating the outflow timescale, we make admittedly simplistic assumptions, including unimpeded outflow motion along a straight path and with constant velocity.  To validate this adopted timescale we also estimate lower and upper limits for the timescale.  Assuming a typical jet velocity of $100$ \kms \ \citep[i.e.][]{Arc10} and an average outflow length of 0.15 pc (as above), a lower limit on the age of the outflow is  $1.5\times10^3$ yr.  As an upper limit for the lifetime of most sources that drive outflows in this region, we adopt 0.5 Myr which is the lifetime for the Class I stage reported by \citet{Eva09}.  All identified outflows in our map are driven by Class 0 objects, except for SVS 13A and IRAS 2B which are likely very young Class I objects.  Given the span of nearly two orders of magnitude for the lower and upper limits for the timescales stated here, we consider the mean dynamical time of $t_{dyn}\sim5\times10^4$ years to be a reasonable median value, acknowledging the uncertainty in this value as cautioned by \citet{Arc10}, and we use this as the outflow timescale, $\tau_{out}$, in the following estimate of outflow luminosity.

We calculate the outflow luminosity to be $L_{out}=E_{out}/\tau_{out}=5\times10^{32}$ erg s$^{-1}$ (or $L_{out}=2\times10^{33}$ erg s$^{-1}$, with $\xi=57.3\dg$).  Turbulent dissipation rate is given by $L_{turb} = E_{turb}/t_{diss}$, where $t_{diss}=5.7\times10^5$ yr is the energy dissipation time for NGC 1333 from \citet{Arc10}.  For the region we mapped, we calculate $L_{turb}=10^{32}$ erg s$^{-1}$, yielding a ratio $r_L=L_{out}/L_{turb}\sim5$.  As is the case for the majority of outflow regions studied by \citet{Arc10}, the ratio $r_L$ is greater than one, signifying that in this region the outflows have more than enough power to be an important agent for the maintenance of turbulence.

These energy diagnostics pertain to only the region we mapped, which is the central part of the cluster.  Even if outflows indeed disrupt the central region, we cannot conclude that they disrupt the entire cluster, for which we would need to map a larger area.  However, our results imply that the star formation activity will be seriously affected in the central part of the cluster, and if this area is disrupted and dispersed, star formation could be halted in the most active and densest region within the cloud.

\subsection{Protostellar sources within the region}

Within our mapped region of $\sim0.23$ pc$^2$, we detect nine continuum sources, and a total of twenty-two outflow lobes or features.  The continuum sources have masses of $\sim0.1-2$ \msun\ (see Table \ref{tab:contsources}, and discussion in \S\ref{sec:cont}).  Of these nine continuum sources, six have associated outflows. In regard to the twenty-two outflow lobes and features, these are driven by 12 distinct sources, of which 8 are previously identified protostellar sources, and 4 are driven by unidentified sources (i.e. outflow candidates C1, C2, C3 and C4).  Within the same region are 55 YSOs and 3 starless cores (see Table \ref{tab:ysos}).  The YSOs include eleven Class 0 sources, eleven Class I sources, four flat-SED sources, twenty-five Class II sources, and four Class III sources.  Of these, 31 YSOs (seven Class 0, four Class I, three flat-SED, sixteen Class II, and one Class III) and one starless core reside within the highest sensitivity region (inner $\sim6\arcmin\times6\arcmin$, see Figure \ref{fig:gut08}). 

Considering the relatively small number of sources in our region, we do not aim to present statistically significant trends, but rather a qualitative discussion of sources and ages within the region in the context of clustered star formation.  Of the 55 YSOs in the entire map, we identify outflows associated with 6 (of 11) Class 0 sources and 2 (of 11) Class I sources.  We find no outflow-driving sources beyond Class I.  Furthermore, we note that the two outflow-driving Class I sources that we identify are likely very young Class I sources.  We also note that seven Class I sources (of the 11 considered here) and four Class 0 sources (SK 1, SK 18, IRAS 4C, IRAS 4D) lie within the outer region of our map where we have lower sensitivity, and therefore we would need to extend the mapped region to better investigate the outflow emission associated with these sources.  In particular, the candidate outflow lobes C3 and C4 appear to originate near the Class I source IRAS 7.  If we consider only the inner $6\arcmin\times6\arcmin$ region with best sensitivity, 2 out of 4 (or 50\%) of the Class I, and 5 out of 7 (or 70\%) of the Class 0 sources in this region drive identified outflows.  

In a study of the CO data in the FCRAO Taurus Molecular Cloud survey, \citet{Nar12} found that 75\% of Class 0 sources, 30\% of Class I sources and 12\% of flat-SED sources known within the Taurus region drive outflows.  Our detection rate within the region of NGC 1333 that we observed with best sensitivity is consistent with that of \citet{Nar12}, especially considering that we mapped a much smaller area (and our sample is smaller). 

We propose several explanations for why we associate outflows with a greater percentage of younger rather than older objects.  First, outflow morphology and collimation changes with time, with the younger outflows more collimated, and the older outflows having wider opening angles \citep{Arc07}.  While we expect that all Class 0 and Class I sources should drive outflows, it is possible that within the lifetime of a Class I source ($\sim 0.5$ Myr), an outflow's morphology becomes diffuse enough as to become much less feasible to associate outflow emission with its driving source.  Proper motions of outflow sources \citep[e.g.][]{Bal01,Goo04} may also make it harder to identify older sources with their respective outflows. 

Alternatively, if outflows are episodic by nature, then perhaps the episodes are more intermittent at later evolutionary stages, and the emission from outflow events has had time to propagate farther from their driving sources (and to slow down), making association of outflows more difficult for these sources.  Based on the sources in the region mapped with highest sensitivity, we associate outflows with 5/7 Class 0 sources but only 2/4 Class I sources.  Assuming that all Class 0 and Class I sources indeed drive outflows, and given a Class 0 lifetime of 0.17 Myr \citep{Eno09} and Class I lifetime of 0.5 Myr \citep{Eva09}, we crudely estimate that outflows are ``on" an average of about 50-70\% of the time during the Class 0 and I stages, or about $\sim0.1$\ Myr and $\sim0.25$\ Myr, respectively.  This is consistent with the results of \citet{Nar12} which indicate that outflows are relatively short-lived based on the non-detection of outflows in majority of the Class I and flat-SED sources in their map.  

Interestingly, one of the Class 0 sources in our map for which we do not identify an outflow (IRAS 4D) has values of $\alpha$ and $T_{bol}$ from \citet{Eva09} which indicate that it is a very young Class 0 source, similar to SK 1 which drives an outflow with very ``knotty'', likely episodic emission. If in fact IRAS 4D and SK 1 are younger Class 0 sources in our map, and considering the average time in which outflows are ``on'' during each phase, then we suggest that the beginning of the Class 0 phase drives short bursts of outflow, followed by progressively longer-duration outflows throughout the Class 0 and Class I stages.  Based on these arguments, our observations indicate that the relative ages of regions may be correlated with the fraction of Class 0 and Class I sources driving outflows, particularly more collimated outflows.  We also acknowledge that the issue of associating outflows with their driving sources is exacerbated in clustered star-forming regions like NGC 1333, where emission is already confused among multiple coincident outflows. 

In summary, we detect 9 continuum sources, and outflows are driven by 12 distinct sources in our map.  This corresponds to 35 continuum sources and about 50 outflow-driving sources per square parsec.  It is also possible that some sources outside of the mapped region are driving outflows which flow into this region, or conversely that sources near the edges of the maps drive outflows outside of the region, so this value is only approximate and will likely become more significant when a larger region is mapped.  Nonetheless, these diagnostics confirm that this is a very active region of star-formation.   In addition, there is at least one starless core (HRF51) \citep{Hat07a} in the region we mapped (with highest sensitivity), which may also form one or more protostars that will drive outflows, adding to the volume occupied by outflows as well as momentum and energy of the region over time.

\subsection{Mean outflow properties, compared with models}
Average mass, momentum and energy of all outflow lobes and features in our map are 0.3 \msun, $0.9$ \msun \kms \ and $3 \times10^{43}$ erg, respectively. Considering just bipolar outflows, the average mass, momentum and energy of outflows (of both lobes) are 0.7 \msun, $2$ \msun \kms \ and $8 \times10^{43}$ erg, respectively.  \citet{Kne00} calculate a total of  1 \msun, $10$ \msun \kms \ and $1 \times10^{45}$ erg in outflows, assuming an arbitrary mean outflow inclination of 45\dg.  They identify 10 outflow-exciting sources in their  $\sim63$ square arcmin map, corresponding to mean outflow mass, moment and energy of approximately 0.1 \msun, $1$ \msun \kms \ and $1 \times10^{43}$ erg, respectively.  In comparison, correcting for an inclination of 45\dg, we measure mean outflow mass, moment and energy of 0.7 \msun, $3$ \msun \kms \ and $16 \times10^{43}$ erg, respectively, larger than the values presented by \citet{Kne00}.  The difference likely arises from the different method used to estimate mass by \citet{Kne00}. We believe that our method, which uses $^{13}$CO to correct for the velocity dependent opacity of the $^{12}$CO line (see \S \ref{sec:mass}) results in more reliable mass estimates \citep[e.g.][]{Off11}.

Some models of outflows, and in particular outflow-driven turbulence, parametrize outflow strength assuming that momentum is proportional to stellar mass \citep[e.g.][]{Mat00,Nak07}.  \citet{Mat00} calculate wind momentum as $p_w=f_w v_w m_*$, where the fraction $f_w$ of the final mass  $m_*$ of the protostar driving the wind gives the wind mass $m_w$, and  $v_w$ is wind velocity.  They justify the assumption that neither $f_w$ nor $v_w$ depends on mass, based on the X-wind model of \citet{Shu94}, and values of $f_w=1/3$ and $f_w=1/10$ correspond to the X-wind and disk wind theories, respectively.  In their simulations, \citet{Nak07} adopt a value of $f=0.5$ in their standard models, also exploring values of $f=0.25-0.75$.  

For an average YSO mass of 0.5\ \msun \ \citep{Eva09} and inclination corrected (with $\xi=57.3\dg$) mean bipolar outflow momentum of  $4$ \msun \kms, we estimate $f\sim 0.08$, assuming conservation of momentum between wind and molecular outflow (basically, entrained cloud gas) and $v_w=100$ \kms.  Uncertainty in $f$ may be due to our choices of $v_w$ and excitation temperature, discussed in \S \ref{sec:mass}, each resulting in uncertainties in the momentum estimate of a factor of about two. We assume that all of the protostellar wind's momentum is used to drive the observed molecular outflow, which should be the case due to conservation of momentum, but perhaps the molecular outflow does not fully trace the wind's momentum.  \citet{Nak07} use higher values of $f$ in their simulations, although they acknowledge that their value for $p_w$ is likely biased towards high velocities and that weaker outflows are difficult to study in detail, whereas our estimates may be biased towards lower values because of the velocity coverage of our observations, reaching only outflow velocities of $|v_{out}|\lesssim10$ \kms, and we also do not know the correction for outflow inclination. 

Of course, our observations are of one particular young region with an age of about 1 Myr (and where outflow sources are in the Class 0 and Class I stages), and therefore it is likely that more momentum will be imparted by outflows as the sources in this region evolve.  In order to quantify the effect of this, we assume that outflows drive constant momentum during the Class 0 and Class I stages, and that these stages have durations of 0.2 Myr and 0.5 Myr, respectively  \citep{Eno09,Eva09}.  This leads us to estimate that the outflows we observe here may last for about two to three times the current outflow driving time. Consequently, the total momentum from outflows may be up to three times the current momentum in this region, and we suggest an upper estimate of $f\sim 3\times0.08=0.24$.  

Since we and others find average outflow momentum in NGC 1333 to be on the lower end of the outflow momentum implemented in models of turbulence injection in protostellar clusters, we suggest that lower values of $f$ are more consistent with observations.  Thus, we propose that models should investigate values of $f$ in the range $\sim0.1-0.25$, in order to understand the impact of outflows during Class 0/I.  Higher values of $f$, such as those with $f=0.25-0.75$ in the simulations of \citet{Nak07}, do not seem to be supported based on our observations.  We also suggest that simulations implement episodic outflows (discussed in the previous section), rather than instantaneous or constant injection of momentum, based on our results.

%%%%%%%%%%%%%%%%%%%%%%%%%%%%%%%%%%%%%%%%%%%%%%%%%
\section{Summary} \label{sec:summary}

We observed the protostellar region NGC 1333 using CARMA, and combining with single dish observations from FCRAO, we present mosaic maps that are sensitive on scales from $\sim 5\as$ to the map size $\sim 7^\prime$ ($0.006$ to $0.5$ pc at a distance of 235 pc).  In this paper we focus on the $^{12}$CO emission which traces cool molecular outflows, and we use simultaneous observations of $^{13}$CO in order to obtain reliable mass estimates.  We identify a total of 22 outflow lobes and features, of which 18 comprise 9 identified bipolar outflows, while 4 outflow features have no identified bipolar counter-part, likely because these outflows extend beyond the edges of our map.  We detect 9 continuum sources in our map, of which 6 are responsible for driving a total of 14 outflow lobes that we identify, with one of these sources driving perpendicular outflows.

The spatial ($\sim 5\as$) and spectral (0.3 \kms) resolution of our observations facilitate our detailed description of outflow morphologies in the region, including identifying outflow-driving sources where possible, as well as evidence for interaction between the outflows in this dense clustered region.  We detect CO emission in 70\% of the $49$ square arcmin (0.23 pc$^2$) area of the map, and we estimate an outflow volume filling factor of $2-6$\%.  We calculate a total of 6 \msun \ of gas associated with identified outflows and outflow candidates, compared with a cloud mass of 140 \msun \ corresponding to the region we mapped. 

We calculate kinematics of the individual identified outflows within the region, as well as total outflow momentum and energy, which are $19$ \msun \kms \ and $7.2\times10^{44}$ erg, respectively ($35$ \msun \kms \ and $2.5\times10^{45}$ erg, respectively, assuming an average inclination angle with respect to the line of sight of $57.3\dg$ for all outflows).  Comparing with gravitational binding energy and turbulent energy for the same region, we find that outflow energy is comparable to turbulent energy, and about one-third of the gravitational potential energy, assuming an average outflow inclination angle of $\xi=57.3\dg$.  We also calculate the ratio of outflow to turbulent luminosity, which is greater than one.  These diagnostics lead us to suggest that outflows act as an important agent for the maintenance of turbulence in this region.  If enough mass is ejected by outflows in excess of the cloud escape velocity, the gravitational binding energy of the cloud could diminish to the point that outflow energetics substantially disrupt the region of the cluster which we mapped, where most stars are forming.  

We also investigate characteristics of protostars and outflows in the region mapped, where about 30\% of the YSOs within all of NGC 1333 reside.  We identify outflows associated with 5 out of 7 Class 0 and 2 out of 4 Class I sources within the region mapped with the highest sensitivity.  We suggest that while we expect that all Class 0 and Class I sources should drive outflows, with our observations we are not able to associate outflows with all of these sources due to (a) difficulty identifying outflow features as they become older and less-collimated, and/or (b) more sporadic outflow episodicity at later stages.  The challenge of identifying outflow features is only augmented in a clustered region like the prototypical NGC 1333, but this is a pertinent issue considering most stars form in embedded cluster environments.

Finally, we make several suggestions of parameters that are important for simulations of clustered star-forming regions.  In the region we mapped, the average mass, momentum and energy of outflow lobes and features are 0.3 \msun, $0.9$ \msun \kms \ and $3 \times10^{43}$ erg, respectively (with no inclination correction).  Our results support a lower value of $f_w$, which parametrizes outflow strength, than has been implemented in previous simulations. In other words, about $10-25$\% of the final mass of a protostar comprises the protostellar wind, which in turn drives momentum in the cluster environment.   Models should take these parameters into account when investigating the impact of outflows during the first $\sim 0.5$ Myr of protostellar evolution.  Our results pertain specifically to the central, most active region of clustered star-formation, and we suggest that the dynamics of this region are critical in understanding the broader impact, and particularly the dispersal, by outflows within the larger cluster environment.

%%%%%%%%%%%%%%%%%%%%%%%%%%%%%%%%%%%%%%%%%%%%%%%%%
%%%%%%%%%%%%%%%%%%%%%%%%%%%%%%%%%%%%%%%%%%%%%%%%%
%% The \notetoeditor{TEXT} command allows the author to communicate
%% information to the copy editor.  This information will appear as a
%% footnote on the printed copy for the manuscript style file.  Nothing will
%% appear on the printed copy if the preprint or
%% preprint2 style files are used.


%% Included in this acknowledgments section are examples of the
%% AASTeX hypertext markup commands. Use \url without the optional [HREF]
%% argument when you want to print the url directly in the text. Otherwise,
%% use either \url or \anchor, with the HREF as the first argument and the
%% text to be printed in the second.

\acknowledgments

We would like to thank an anonymous referee for suggestions that strengthen the quality of the paper.  ALP would like to thank J. Carpenter for help with data reduction, and M. Pound for advice on combining interferometer and single dish maps. This material is based upon work supported under a National Science Foundation Graduate Research Fellowship, and support from the US Student Program of Fulbright Chile.  This project was funded by the NSF under grant AST-0845619 to HGA.  DM gratefully acknowledges support from CONICYT project BASAL PFB-06.  Support for CARMA construction was derived from the Gordon and Betty Moore Foundation, the Kenneth T. and Eileen L. Norris Foundation, the James S. McDonnell Foundation, the Associates of the California Institute of Technology, the University of Chicago, the states of California, Illinois, and Maryland, and the NSF. Ongoing CARMA development and operations are supported by the NSF under a cooperative agreement, and by the CARMA partner universities.  The National Radio Astronomy Observatory is a facility of the National Science Foundation operated under cooperative agreement by Associated Universities, Inc.


%% in parentheses, after the keyword, but they will not be verified.

{\it Facilities:} \facility{CARMA}, \facility{FCRAO}.

\bibliography{ms}{}
\bibliographystyle{apj}

%%%%%%%%%%%%%%%%%%%%%%%%%%%%%%%%%%%%%%%%%%%%%%%%%
%tables
\clearpage
%%%%%%%%%%%%%%%%%%%%%%%%%%%%%%%%%%%%%%%%%%%%%%%%%

%%%%%%%%%%%%%%%%%%%%%%%%%%%%%%%%%%%%%%%%%%%%%%%%%
%figures
\clearpage
%%%%%%%%%%%%%%%%%%%%%%%%%%%%%%%%%%%%%%%%%%%%%%%%%

%% The following command ends your manuscript. LaTeX will ignore any text

\end{document}